\documentstyle[11pt,emlines]{article} \textwidth=17cm
\begin{document}

\begin{center} {\Large P.N.Lebedev Physical Institute Russian AS} \end{center}
\begin{center} {\Large Nuclear Physics and Astrophysics Division} \end{center}
\begin{center} {\Large High Energy Physic Department} \end{center}
\begin{center} -------------------------------------------- \end{center}

\vskip 5cm \begin{center} {\large Preprint No 35} \end{center} \vskip
2cm \begin{center} {\Large \bf Long wavelength broadband sources of
coherent radiation} \end{center} \vskip 1cm \begin{center}{Yu.Shibata }
\end{center} \begin{center} {Research Institute for Scientific
Measurements, Tohoku University, Japan} \end{center} \begin{center}
E.G.Bessonov \end{center} \begin{center} Lebedev Physical Institute of
the Academy of Sciences, Moscow, Russia \end{center} \vskip 7cm
\begin{center} Moscow 1996 \end{center}
\vfill\eject\empty \hskip 5cm \empty

\footskip 10mm
\vskip 2cm \setcounter {page} {2}

                        \begin{abstract}
Some considerations of long wavelength and broadband radiation sources based
on the emission of the coherent radiation by a train of short relativistic
electron bunches moving in an open resonator along an arc-like or undulator
trajectories and some new versions of the transition radiation sources and
long wavelength sources based on storage rings are presented.    \end{abstract}

\newpage
\begin{center} {\Large \bf Contents.} \end{center}

{\large \bf 1. Introduction}  \hfill  3\\

{\large \bf 2. Particle radiation in the external fields and in media}
\hfill 4\\

{\large \bf 3. Stimulation of the energy lost of particles in the
external}\\
\empty \hspace*{10mm} {\large \bf fields by an open resonator} \hfill
7\\

{\large \bf 4. Prebunched free-electron laser based on an undulator
and}\\ \empty \hspace*{10mm} {\large \bf on an open resonator} \hfill
10\\

{\large \bf 5. Transition radiation sources } \hfill 13\\

\empty \hskip 5mm {\large \bf 5.1. Introduction to the theory of the
transition radiation} \hfill 13\\

\empty \hskip 5mm {\large \bf 5.2. The transition radiation of
particles crossing a series of }\\
\empty \hspace*{20mm} {\large \bf two developed at the angle 45$^o$
mirrors} \hfill 15\\

\empty \hskip 5mm {\large \bf 5.3. Transition radiation of particles
passing through }                  \\
\empty  \hspace*{20mm} {\large \bf a system of mirrors} \hfill 18\\

\empty \hskip 5mm {\large \bf 5.4. Stimulated transition radiation}
\hfill 21 \\

{\large \bf 6. The long wavelength sources based on the storage rings}
\hfill 21\\

{\large \bf 7. On the choice of the scheme of the long wavelength
broadband }                          \\
\empty \hspace*{10mm} {\large \bf source of coherent radiation} \hfill
23\\

{\large \bf 8. Conclusion} \hfill 24\\

{\large \bf 9. Appendix 1} \hfill 25\\

{\large \bf 10. Appendix 2} \hfill 26 \\

{\large \bf 10. Appendix 3} \hfill 27 \\

{\large \bf 11. References} \hfill 29 \\

\vfill \eject

                     \section{Introduction.}

Many fields of science are in need of the long wavelength and broadband
radiation sources with a continuous spectrum. The blackbody sources
and sources of spontaneous incoherent synchrotron and undulator radiation
based on storage rings of the energy $\sim $ 1 GeV and higher are used in
this case. Such sources are effective in optical and harder regions of
spectra. In IR and longer wavelength regions of spectra only monochromatic
radiation sources like lasers (masers), free-electron lasers, traveling-wave
tubes, klystrons and so on are effective.

The problem of high power long wavelength and at the same time broadband
sources with continues spectrum is open now. One of the possible ways to
solve this problem is the way using the coherent radiation of short bunches
of relativistic electrons passing through the external fields or through
or near material bodies (Cherenkov, transition, diffraction radiation).

Effective generation of coherent radiation is possible when the electron
bunch-length is less or comparable with the emitting wavelength. It means
that the spectrum of the broadband sources is limited from the
short wavelength region by the electron bunch length.

At present all accelerators and special bunching systems produce periodical
trains of bunches under a definite bunching frequency. It means that all
generators based on such beams will emit line spectrum radiation at bunching
frequency and their harmonics. Line nature of the spectrum can be displayed
in the long wavelength region. In this case to generate spectrum similar to
continuous one the distance between bunches must be much (some orders)
greater then the wavelengths of the emitted radiation.

Modern high energy linear accelerators using electron guns with thermal
cathodes and bunchers can produce now electron bunches of the length $\sim $
1mm and less. New technology connected with electron guns based on laser
photocathodes permits to produce shorter bunches. Bunchers based on
undulators and lasers permit to produce a long train of very short
microbunches at the bunching frequency of the lasers but such train of
microbunches could not be used in the long wavelength and broadband sources
of the coherent radiation with a continuous spectrum.

In this paper we search different schemes of coherent radiation sources.

\section{Particle radiation in the external fields and in media.}

The electric and magnetic field strengths of a non-uniformly moving charged
particle are determined by expressions
                      \begin{equation}   
   \vec E(t) = \vec E^c(t) + \vec E^r(t),\hskip 15mm \vec H = [\vec n \vec E],
                     \end{equation}
where
          $$\vec E^c(t) = {e(1-\beta ^2)(\vec n-\vec \beta)
          \over R^2(1-\vec n\vec \beta)^3}|_{t^{'}}, \hskip 15mm
          \vec E^r(t) = {e[\vec n[(\vec n-\vec \beta )\dot {\vec \beta
          }]]\over cR(1-\vec n\vec \beta)^3}|_{t^{'}},$$
e, c$\vec \beta$, c$\dot {\vec \beta }$ are the charge, velocity, and
acceleration of the particle, $\vec n$ is the unit vector directed from the
particle to the observation point, $R$ is the distance from the particle
to the observation point, and $t$ is the time of observation \cite{landau},
\cite{jackson}. On the right-hand side of the expressions (1), $\vec \beta$,
$\dot {\vec \beta }$, $\vec n$ and $R$ must be taken at the earlier time
$t^{'} = t - {R(t^{'})/c}$.

The first term in (1) describes the sharply decreasing Coulomb field of the
particle, while the second describes the electromagnetic field radiated by
the particle. When $\dot {\vec \beta } = 0$ then the emission of the
electromagnetic waves is absent. The energy of the electromagnetic radiation
emitted by the particles can be taken from the kinetic energy of the
particles or from the energy of the extraneous force sources.

The power emitted by a particle is determined by the instantaneous values of
the velocity and acceleration of the particle:
                      \begin{equation}   
P = {d\varepsilon \over dt} = {2e^2\over 3c^3}(|\dot {\vec \beta }|^2 - (\vec
\beta \dot {\vec \beta })^2)\gamma ^6,
                      \end{equation}
where $\varepsilon $ is the energy, $\gamma = \varepsilon /mc^2$ is the
relativistic factor, and $m$ is the mass of the particle. The angular
distribution of the power emitted by the particle
                      \begin{equation}   
           {dP\over do} = {c\over 4\pi}|E^r|^2R_o^2 = {e^2\over 4\pi c^3}\{{2(
          \vec n \dot {\vec \beta })^2\over c(1-\vec n\vec \beta)^5} + {|\dot
           {\vec \beta}|^2\over (1-\vec n\vec \beta)^4} - {(\vec n \dot {\vec
           \beta })^2\over (1-\vec n\vec \beta)^6 \gamma ^2}\}  \end{equation}
at the distances $R_o$ much higher then the dimensions of the region where
the radiation is emitted (acceleration is equal zero). In this case we can
suppose $R = R_o =const.$

The spectral-angular and polarized properties of the emitted radiation are
defined by Fourrier's transform $\vec E_{\omega}^r$ of the electric field
strength vector $\vec E^r(t)$. Further we will use the Fourrier's transform of
the form
           \begin{equation} 
           \vec {E_\omega } = {1\over 2\pi}\int _{- \infty}^{+ \infty}\vec
           E^r(t) e^{i\omega t}dt = {1\over 2\pi}\int _{t_1}^{t_2}\vec E^r(t)
           e^{i\omega t}dt, \end{equation}
where the time $t_1$ corresponds to the time of $t_1^{'}$ of the particle
entering in and the time $t_2$ corresponds to the time of $t_2^{'}$ of the
particle exit from the external field. They are determined by the conditions
$\dot {\vec \beta }(t < t^{'}_1) = 0$, $\dot {\vec \beta }(t > t_2^{'}) = 0$,
$|\dot {\vec \beta}(t_1^{'} < t < t_2^{'})| \geq 0$, that is by the part of
the trajectory inside which the acceleration of the particle differ from zero.

Thus, for instance the spectral-angular distribution and the spectral density
of the energy emitted by a particle and passing through the element of the
solid angle $do = dS/R_o^2$ or the area $dS$ at the observation point are
determined by the expression
                      \begin{equation}   
           {\partial ^2\varepsilon \over \partial \omega \partial o} =
          R_o^2{\partial ^2\varepsilon \over \partial \omega \partial S} =
           cR_o^2|\vec E_{\omega}|^2.                   \end{equation}

The Fourrier's transform of the electromagnetic field strength at zero
frequency $\omega = 0$ can be presented in the form $\vec
E_{\omega}|_{\omega = 0} = (1/2\pi )\vec I$, where
          \begin{equation} 
          \vec I = {e\over cR_o}[\vec n [\vec n({\vec \beta _2\over
          1-\vec n\vec \beta _2} - {\vec \beta _1\over 1 - \vec n \vec
          \beta _1})]], \end{equation}
is the strange parameter of the
emitted wave, subscripts $1,2$ relate to initial and final electron
velocities or the same to the moments $t_1$, $t_2$. By definition
"Strange electro-magnetic Waves" (SW) are waves, whose electric field
vector $\vec E(t)$ satisfies the condition $\vec I = \int
_{-\infty}^{+\infty}\vec E(t)dt \ne 0$. Such waves and the strange
parameter $\vec I$ were introduced in \cite{bes1a}, \cite{bes1b} to
describe some of the properties of the emitted radiation. The value
$\vec I \ne 0$ when $\vec \beta _1 \ne \vec \beta _2$. It means that SW
are emitted only in the case of the unlimited trajectories of the
particles (either $\vec \beta _1 \ne 0$ or $\vec \beta _2 \ne 0$ or
both of them differ from zero ). In particular, waves for which the
components have the form of unipolar pulses are strange one. SW
transfer to a charged particle at rest a momentum linear in $\vec E$ and
directed perpendicular to the direction of their propagation \footnote[1]{When
$|\vec I \ne 0|$ then spectral intensity (4) is not equal zero up to zero
frequencies or up to the wavelengths $\lambda \to \infty$. It means that in
this case the part of the emitted radiation with the wavelengths $\lambda >
R_o$ is not in the far (wave) zone and we can't consider this part of waves
as plane waves in this region.}.

Notice that the Maxwell equations allow the arbitrary and hence unipolar
pulses of the electromagnetic waves. However this is necessary but not
sufficient condition of the existing of such waves. It is necessary to
have the conditions for the emission of such waves. The sufficient
condition for the emission of SW is  $\vec \beta _1 \ne \vec \beta _2$.

Notice also that this conclusion is valid when the distance $R$ in (1) is much
greater then the dimensions of the emission region \cite{landau},
\cite{jackson}. In this case we can consider that the radiation is emitted
from small (point-like in comparison with $R$) region and propagate under
the same direction defined by the constant unite vector $\vec n$ and at the
constant distance $R = R_o$ \footnote[2]{Only this case is under consideration
in the textbooks when spectral-angular distributions of the emitted radiation
are investigated.}. However the conditions can be realized when the particle
pass through two regions $A$ and $B$ and the distances between this regions
and observation point are near the same. Under such conditions particle can
emit two waves with strange parameters $\vec I_{A} \ne 0$ and $\vec I_{B} \ne
0$ when the initial and final velocities of the particle are zero. These
waves will come to the observer under different angles (different unite
vectors $\vec n_{A}$, $\vec n_{B}$) and with different values of the vectors
$|\vec I_{A}|$, $|\vec I_{B}|$. They can be selected this place by splits or
concave mirrors \cite{bes1a}, \cite{bes1b}. Such waves will have increased
intensity in the
long wavelength region and can be used this place or after separation at more
distant places. When the distance from each region to the observation point
is much greater then the distance between regions then in this case $\vec
I_{A} = - \vec I_{B}$, $\vec I = 0$ and the efficient conditions for the
long wavelength radiation are absent. {\it One of the important elements in the
projects of the long wavelength radiation sources based on the relativistic
electron beams can be the selection elements}.

The fruitful idea of the formation zone was introduced in \cite{ginzburg},
\cite{bolot}. {\it But this idea works well under conditions when the
distance from all emitting regions to the observer is much larger then
distances between these regions and theirs dimensions and when there is no
focusing and other elements between these regions  capable to destroy the
interference between the emitted wavepackets}. The total spectrum of
the emitted radiation passing through the closed surface including regions
$A$ and $B$ depends on the shape of the surface and it's distance from the
emitting regions.

Electromagnetic radiation (electromagnetic waves) can be emitted by charged
particles in free space {\it only under conditions of nonuniform
(accelerated)} motion of the particles. This statement fallow from the
expressions (1),(2) describing the electromagnetic field strengths and power
of the emitted radiation. When the uniformly moving charged particle
(projectile particle) pass through a transparent medium and near or through a
material body or a system of bodies it emit Cherenkov and transition
(diffraction) radiation respectively \cite{ginzburg}. One
speaks that in this case the radiation is emitted by uniformly moving
particle. However this is spoken for simplicity and convenience. In reality
in this case the energy of the emitted radiation is taken from the extraneous
forces source. This source transform the energy to the projectile particle
support it on the constant level and through this particle transform it to
particles of the medium. The particles of the medium undergo the acceleration
and hence emit radiation and decelerate projectile particle with the force
equal to and directed backward to the extraneous force. No waves go away from
projectile particle. If the extraneous forces are absent, then the heavy
projectile particle will be decelerated by fields induced by media particles
but the electric field strength and hence the energy of the wave emitted by
the projectile particle will be much less then the field strength and the
energy emitted by particles of media. In this case the total radiation will
be emitted by the accelerated particles of media and by the accelerated
projectile particle. It means that according to these common positions all
known kinds of radiation can be reduced to a radiation of accelerated
particles.

Usually the electron beams of linear accelerators consist of trains
of bunches. Each train (impulse) is repeated with some industrial frequency
$f_o$ and each bunch in the train is repeated with some bunching frequency
$f_b$. Each bunch in it's turn can consist of a small microbunches following
under theirs microbunching frequencies $f_{mb}$ \footnote[3]{The frequency
$f_o$ is the switching on frequency of the generator of the linear accelerator
system, the frequency $f_b$ is the frequency of the generator (klystron and so
on)  and $f_{mb}$ is the frequency of the microbunching laser wave of a
special bunching system (based on undulator and other devices).}.
Such electron beams in arbitrary external fields or in resonators and any
other systems emit electromagnetic radiation in the form of the wavepackets
with the same repetition frequencies. It means that according to Fourrier's
transform theorems the frequencies of the emitted waves in such systems will
have line spectrum of the frequencies $f_{mb}$, $f_b$, $f_o$, theirs harmonics
and frequency combinations of the form \footnote[4]{This result can be
received by analogy with the derivation of the expression (26) (see below).}
            \begin{equation}  
          f_{nmk} = nf_{mb} \pm mf_b \pm kf_o,
            \end{equation}
where n,m,k are whole numbers. Usually $f_o \ll f_b \ll f_{mb}$. The frequency
$f_o$ ($f_o \sim $100 Hz) is many ($\sim $7) orders less then $f_b$ ($f_b \sim
$1 GHz) and we can neglect the splitting of the frequencies by the value $f_o$.

In the examples below we will deal with frequencies $f_b$ and $f_o$ of the
beam bunched in the accelerator and we will neglect the frequency splitting
of the value $f_o$. A single bunch in this case will emit radiation with some
continuous spectrum. The spectrum of a large number  of bunches $N_b \gg 1$
will look like line spectrum with envelope of the continuous spectrum. The
bandwidth of the lines will be determined by the number of the bunches $N_b$
and the harmonic number $m$: $\Delta f/f \simeq 1/mN_b$. The distance between
lines $f_b$ can be resolved in the long wavelength region ($\lambda
\sim 1-10^{-2}$ cm, $c/f_b \sim 10$ cm).

When high quality resonator is used with some natural mode frequencies $f_r$
then only modes with frequencies equal or near to bunching frequencies or
theirs harmonics will be exited.

       \section{Stimulation of the energy lost of particles in the
       external fields by an open resonator.}

A scheme of the source based on coherent radiation of bunched relativistic
electron beam in external fields and intensified by an opened resonator is
presented in the Fig.1. On this Figure the external fields are presented by an
ordinary banding magnet.

Electron beam from a linear accelerator enter the vacuum chamber installed in
a bending magnet at the angle $\alpha _1$ to the axis $y$, pass trough a thin
flat metal mirror $M1$, go along an arc through magnetic field which is built
up from zero to a maximum value $B_m$ at the length of the order of the gap of
the bending magnet and then exit from the resonator at the angle $\alpha _2$
to the axis $y$ through the hole (or through a thin foil installed in the
hole) of the second spherical concave mirror $M2$ of the resonator. Some hole
(window) in resonator mirrors or small mobile mirror M3 inserted in
the resonator can be used to extract the stored radiation from resonator.

Electron bunches in the magnetic field will emit radiation in the form of a
train of short wave packets. This wave packets will be stored in the cavity
for some time determined by losses in the cavity (quality of resonator). If
the bunching frequency is equal to the oscillating frequency of the
wavepackets in resonator (when the electron velocity $v$ is near to the light
one and double distance between mirrors is multiple to the distance between
electron bunches) and the quality of resonator is high then the wavepackets
of the emitted radiation will be overlapped. The value of the electric field
strength and the power of the stored wavepackets will be proportional to the
effective number and to square of the effective number of stored wavepackets
(photon bunches) accordingly. This way the resonator will enhance the power
of the emitting radiation.  In this case the conversion of the electron beam
energy to the energy of the electromagnetic radiation will go mainly through
the interaction of the transverse electric fields of the stored wavepackets
with transverse  components of the electron velocities obtained by electrons
in the magnetic field but not through the self-decelerating fields.

The length of the stored wavepackets is much smaller then the length of the
resonator. It means that in this case a high number of the longitudinal modes
will be exited in the resonator.

When the combination of plain mirror and spherical mirror is used, then the
waist of the photon beam will be at the flat mirror and the radius of the
waist will be determined by the expression
          \begin{equation} 
          \sigma _{\gamma 0} = ({d(R-d)\lambda ^2 \over \pi ^2})^{1/4},
          \end{equation}
   \vskip -30mm
\begin{center}
\unitlength=0.40mm
\linethickness{0.8pt}
\begin{picture}(210,150)
\put(-90,50){\line(1,0){340}} 
\put(250,50){\vector(1,0){8}} 
\put(0,10){\line(0,1){80}}
\put(-75,50){\line(1,5){3}}
\put(215,50){\line(-1,5){3}}
\put(200,40){\line(0,1){20}}
\put(200,60){\line(-1,2){9}}
\put(200,40){\line(-1,-2){9}}
\put(191,78){\line(-1,1){30}}
\put(191,22){\line(-1,-1){30}}
\put(10,0){\line(0,1){100}}
\put(140,0){\line(0,1){100}}
\put(-80,67){\line(3,-1){120}}
\put(40,27){\line(5,-1){20}}
\put(60,23){\line(1,0){20}}
\put(80,23){\line(5,1){20}}
\put(100,27){\line(3,1){150}}
\put(10,0){\line(1,0){130}}
\put(10,100){\line(1,0){130}}
\put(250,43){\makebox(0,0)[cc]{$y$}}
\put(-80,58){\makebox(0,0)[cc]{$\alpha _1$}}
\put(-50,65){\makebox(0,0)[cc]{$e^{-}$}}
\put(-46,61){\vector(4,-1){10}}
\put(230,74){\vector(4,1){10}}
\put(73,28){\vector(4,0){10}}
\put(220,58){\makebox(0,0)[cc]{$\alpha _2$}}
\put(215,75){\makebox(0,0)[cc]{$e^{-}$}}
\put(65,30){\makebox(0,0)[cc]{$e^{-}$}}
\put(55,110){\makebox(0,0)[cc]{$BM$}}
\put(80,90){\makebox(0,0)[cc]{$ME$}}
\put(-10,20){\makebox(0,0)[cc]{$M1$}}
\put(220,20){\makebox(0,0)[cc]{$M2$}}
\put(0,15){\line(5,0){100}}
\put(100,15){\line(5,-1){40}}
\put(140,7){\line(2,-1){24}}
\put(0,85){\line(5,0){100}}
\put(100,85){\line(5,1){40}}
\put(140,93){\line(2,1){24}}
\put(170.00,24.00){\makebox(0,0)[cc]{\small M3}}
\put(150.00,26.67){\line(1,-1){10.00}}
\put(154.17,18.33){\vector(-1,-4){10.83}}
\put(150.83,-27.67){\makebox(0,0)[lc]{\small Extracted radiation}}
\put(-50.00,-64.00){\makebox(0,0)[lc]{\small Fig.1: Schematic diagram of the
experimental setup. M1: plain mirror; M2: spherical}}
\put(-23.00,-80.00){\makebox(0,0)[lc]{\small mirror; M3: extracting mirror;
BM: banding magnet; ME: mode envelope.}}
\end{picture}
\end{center}
             \vskip 45mm
where $d$ is the distance between mirrors, $R$ is the radius of the spherical
mirror, $\lambda $ is the wavelength of the emitted radiation
\cite{maitland}.

It follows from (8) that resonator works when $R>d$. Usually one choose $R =
2d$. In this case $\sigma _{\gamma 0} \simeq \sqrt {\lambda d/\pi}$.

The field distribution at the arbitrary point considered will be determined by
the expression

         \begin{equation} 
     \sigma _{\gamma} = \sigma _{\gamma 0}\sqrt {1 + ({y\over l_R})^2},
          \end{equation}
where $l_R = {\pi \sigma _{\gamma 0}^2/\lambda }$ is the Rayleigh length.

The photon beam radius is increased $\sqrt 2$ times and the area of photon
beam $2$ times per the length $l_R$. At the distance $\sigma _{\gamma }$ the
density of photon beam in the transverse direction will decay $e$ times
\cite{maitland}.  \\

           $\underline {Example.}$ Let us $\lambda = 1mm, d = 50cm, R = 1m$.
        In this case the value $\sigma _{\gamma 0} \simeq 1.3$cm, $l_R \simeq
        53.1$cm.\\

The resonator must allow a few tens of transitions, before the radiation is
reduced by the various loss processes (transmission, scattering diffraction,
walk out, etc.) to $e^{-1}$ of its initial intensity. The photon beam losses
in resonator determine the quality of a resonator. Diffraction losses are
determined by Fresnel numbers
                    \begin{equation} 
             N_i^F = {r_i^2\over \lambda d},
            \end{equation}
where $r_i = (r_0, r_m)$, $r_0$ is the hole radius of the resonator mirror,
$r_m$ is the radius of the mirror \cite{maitland}.\\

In the case of TEM$_{00}$ resonator mode the photon beam energy loss per pass
$\Delta W/W \sim 10^{-3}$ when resonator Fresnel number is $N_m$ = 1 and
$\Delta W/W \sim 10^{-2}$ when aperture Fresnel number is $N_0^F = 10^{-3}$
\cite{maitland}. When the photon beam energy loss per pass is high ($\Delta
W/W \geq 0.1$) then it can be approximated by the expression $\Delta W/W
\simeq N_0 \simeq (r_0/\sigma_{\gamma 0})^2$ (the ratio of the areas of the
hole in mirror and photon beam). The same consequences are valid for mirror
inserted in the resonator.

The values $r_m$ and $\sigma _{\gamma o}$ are limited. It means that the
range of the wavelengths of the source according to (8) and condition $N_m >
1$ is limited in the long wavelength region. Closed resonator (cavity) can be
used to move this limit down.

The banding radius of the electron trajectory in the bending magnet $\rho =
{mc^2\gamma /eB_m}$ ($\rho [cm] \simeq 1700\gamma /B_m[Gs]$), where $\gamma =
\varepsilon /mc^2$ is the relativistic factor, $B_m$ is the magnitude of the
magnetic field strength of the banding magnet. The bending angle of the
electron $\Delta \theta = \alpha _2 - \alpha _1 \simeq a/\rho $, where $a$ is
the length of the banding magnet. According to (1) the electric field
strength of the wavepackets emitted by electrons in the bending magnet in the
direction of the axis "y" at large distances (mirror M2 is removed) have
maximum value when the electron velocity is parallel to the axis $y$. It has
zero values when the electron velocity is under the angles $\alpha _1 = -
1/\gamma $ and $\alpha _2 = 1/\gamma $ to the axis $y$. The components of the
electric field strength in this case have the form which looks like single
sign near one semi-period sine form of duration $\Delta t = a/c\gamma ^2$ (see
Fig 2a). The wavepackets of single sign components of the electric field
strengths present strange electromagnetic waves. The spectrum of such
wavepackets stretch up to zero frequencies. In this case the spectral density
of the emitted radiation take on maximum value. This is to be expected that
this conditions will be optimum for this geometry of the experiment.

The form of the time dependence of the electric field strength of radiation
emitted in the case of $\alpha _2 = - \alpha _1 \gg 1$ will correspond to the
synchrotron radiation form of wavepackets (see Fig.2b).

When the banding angle $\alpha _1 \simeq \pm 1/\gamma$ and $\alpha _2 \gg
1/\gamma$ (the case when the electron beam pass by the mirror M2 and the
heating problem of the mirror disappear) then the wavepackets of the
electromagnetic radiation emitted by particles in the direction of the axis
$y$ will present strange electromagnetic waves as well (see Fig. 2c, Fig.2d).
But spectral density of the emitted radiation will be $\sim 4$ times less then
optimum one.

The condition of optimal generation in the banding magnet ($\Delta \theta =
2/\gamma$) does not depend on the electron energy. The optimal value of the
magnetic field strength in this case

           \begin{equation} 
           B_{opt} = {2mc^2\over e\,a}
           \end{equation}
or $B_{opt}[G] = 3400/a[cm]$.
\vskip 6mm

$\underline {Example}.$ When the length of the banding magnet is $a$ = 50 cm
then the value of the optimum magnetic field $B_{opt}$ = 68 G.\\

The transition radiation will be emitted by electron in the flat mirror. This
radiation corresponds to instantaneous start of the electron and its  image
of opposite charge on the mirror surface in the opposite directions
\cite{ginzburg}. The duration of the wavepacket of the transition radiation is
very small ($\sim 10^{-6}$ cm). This wavepacket has single sign form of the
electric field strength and hence present "strange wave radiation"
\cite{bes2}. The signs of the electric field strengths of radiation emitted
in the magnetic field and in mirror are opposite. That is why the modulus of
the strange parameter (6) of the total radiation does not depend on the value
of the magnetic field and defined by (6) under condition $\vec \beta _1 = 0.$

In general case ($H \ne 0$, $\alpha _i$ are arbitrary) particle bunches will
not interact with stored transition radiation and hence will not amplify it
as the
electric field strength of the radiation stored in opened resonator near by
the mirrors surfaces (where the transition radiation is emitted) has not both
longitudinal and tangent components.

When an electron is going along the axis $y$ and enter the mirror M1 and
when the magnetic field is switched off ($\alpha _1 = \alpha _2$ = 0) then
the transition radiation will be stored in the resonator but it will not be
amplified as the stored radiation in opened resonator has only transverse
components of the electric field strengths and the electron velocity has only
longitudinal component \footnote[5]{Closed resonator (cavity) can be excited
through the transition radiation as such resonator have the TE modes with the
longitudinal components of the electric field strengths.}.
                \vskip -20mm
\begin{center}
\hskip 32mm
\unitlength=1.00mm
\linethickness{0.4pt}
\begin{picture}(145.00,98.94)
\put(10.00,70.00){\vector(1,0){55.00}}
\put(35.07,63.87){\vector(0,1){35.07}}
\put(41,97){\makebox(0,0)[cc]{\small $E_{\alpha }$}}
\put(52.67,93.33){\makebox(0,0)[cc]{a}}
\put(52.67,93.33){\circle{4.00}}
\put(65.00,66.67){\makebox(0,0)[cc]{$t$}}
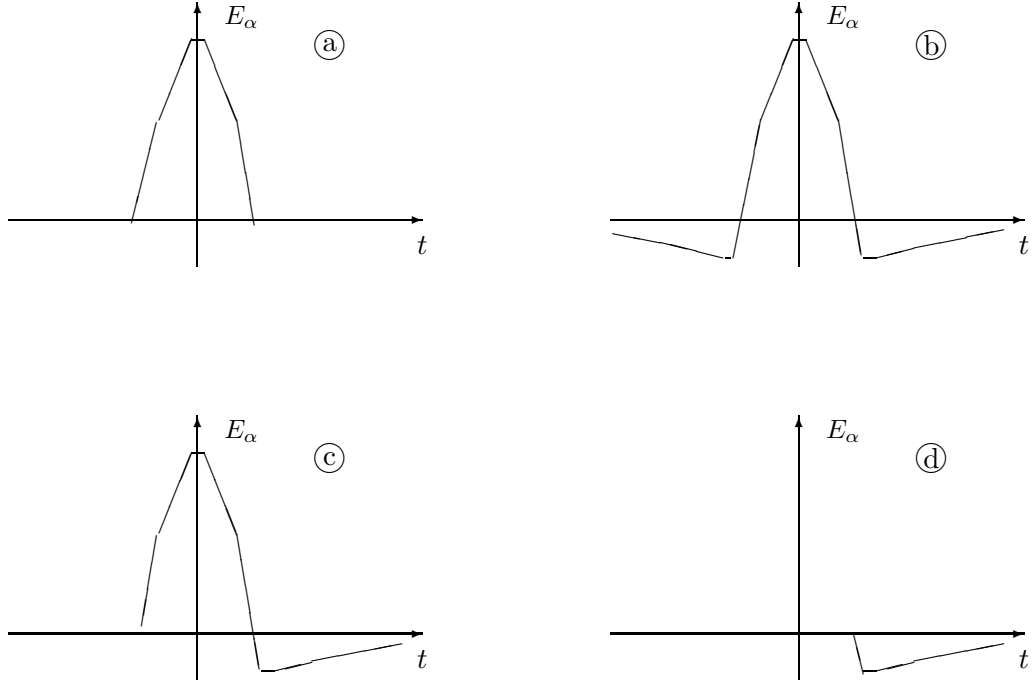
\put(75,-5){\makebox(0,0)[cc]{\small Fig.2: The time dependence of the
electric field strength component $E_{x }$ for $\alpha _1$ =}}
\put(83,-12){\makebox(0,0)[cc]{\small $\alpha _2 = 1/\gamma $ (a); $\alpha _1
= \alpha _2 \gg 1$ (b); $\alpha _1 = -1/\gamma, \alpha _2 \gg 1$ (c);
$\alpha _1 = 1/\gamma , \alpha _2 \gg 1$ (d).}}
\put(30.00,83.33){\line(2,5){4.33}}
\put(34.33,94.00){\line(1,0){1.67}}
\put(36.00,94.00){\line(2,-5){4.33}}
\put(90.00,70.00){\vector(1,0){55.00}}
\put(115.07,63.87){\vector(0,1){35.07}}
\put(106.00,64.93){\line(-1,0){0.67}}
\put(121,97){\makebox(0,0)[cc]{\small $E_{\alpha }$}}
\put(132.67,93.33){\makebox(0,0)[cc]{b}}
\put(132.67,93.33){\circle{4.00}}
\put(145.00,66.67){\makebox(0,0)[cc]{$t$}}
\put(105.00,65.00){\line(-4,1){8.33}}
\put(96.67,67.00){\line(-5,1){6.33}}
\put(123.67,65.00){\line(1,0){1.67}}
\put(125.33,65.00){\line(4,1){5.00}}
\put(130.33,66.33){\line(5,1){12.00}}
\put(106.33,65.00){\line(1,5){3.67}}
\put(110.00,83.33){\line(2,5){4.33}}
\put(114.33,94.00){\line(1,0){1.67}}
\put(116.00,94.00){\line(2,-5){4.33}}
\put(120.33,83.33){\line(1,-6){3.00}}
\put(10.00,15.00){\vector(1,0){55.00}}
\put(35.07,8.87){\vector(0,1){35.07}}
\put(41,42){\makebox(0,0)[cc]{\small $E_{\alpha}$}}
\put(52.67,38.33){\makebox(0,0)[cc]{c}}
\put(52.67,38.33){\circle{4.00}}
\put(65.00,11.67){\makebox(0,0)[cc]{$t$}}
\put(43.67,10.00){\line(1,0){1.67}}
\put(45.33,10.00){\line(4,1){5.00}}
\put(50.33,11.33){\line(5,1){12.00}}
\put(30.00,28.33){\line(2,5){4.33}}
\put(34.33,39.00){\line(1,0){1.67}}
\put(36.00,39.00){\line(2,-5){4.33}}
\put(40.33,28.33){\line(1,-6){3.00}}
\put(90.00,15.00){\vector(1,0){55.00}}
\put(115.07,8.87){\vector(0,1){35.07}}
\put(121,42){\makebox(0,0)[cc]{\small $E_{\alpha }$}}
\put(132.67,38.33){\makebox(0,0)[cc]{d}}
\put(132.67,38.33){\circle{4.00}}
\put(145.00,11.67){\makebox(0,0)[cc]{$t$}}
\put(123.67,10.00){\line(1,0){1.67}}
\put(125.33,10.00){\line(4,1){5.00}}
\put(130.33,11.33){\line(5,1){12.00}}
\put(29.67,83.00){\line(-1,-4){3.33}}
\put(40.33,83.33){\line(1,-6){2.33}}
\put(29.67,28.00){\line(-1,-6){2.00}}
\put(122.33,15.00){\line(1,-4){1.33}}
\end{picture}
\end{center}
                \vskip 15mm

The scheme presented in the Fig.1 is the scheme of the prebunched (parametric)
free-electron laser (fel) based on one banding magnet. Usually a system of
banding magnets of the alternate polarity (undulators) is used in similar
schemes. The theory of the prebunched fel will be presented in the next
section. We propose that this theory in the first approximation will be valid
for short (one or a half period) undulators that is for the case considered
in this section.\\

\section{Prebunched free-electron laser based on an undulator and on an open
resonator.}

A schematic diagram of the parametric (prebunched) free-electron laser based
on  an undulator is presented in the Fig.3.  Some elements of the theory of
such lasers based on open resonators was developed in \cite{al}.
Experimental investigation of the parametric free-electron laser using
cylindrical resonator was made in \cite{ale}. Theory of such schemes does not
completed. Now we present a development of the theory of such lasers for the
case of the helical undulator of the arbitrary length \footnote[6]{The
difference in the intensity of the radiation emitted on the first harmonic in
a plane undulator drawn in Fig.3 and in a helical undulators is negligible
($\sim 30$\%) but the polarization is circular for helical undulators and
linear for plane undulators. Formulas for the particle emission
in a helical undulator are the simplest one.}. This theory will permit to
estimate the power emitted in the previous scheme. A short (one
period) undulator is drawn in the Fig.3. Such undulator with high deflecting
parameter will permit to generate the long wavelength broadband radiation.

The particle trajectory in an arbitrary undulator satisfy the conditions:
$\vec r(t +
mT) = m\vec {\lambda }_u + \vec r(t)$, $\vec \beta(t + T) = \vec \beta (t)$,
where $T$ is the period of the particle oscillations in the undulator, $m =
1,2,3 ... K$, $|\vec {\lambda }_u|$ is the undulator period, vector $\vec
\lambda _u/|\vec \lambda _u|$ is the unit vector along the average velocity
of the particle, $\overline {\vec v} = \vec {\lambda }_u/T$ \cite{alf1}.
The relative electron velocity in the undulator have longitudinal and
transverse components $\vec \beta = \vec \beta _{\parallel} + \vec \beta
_{\perp}$. In helical undulators the values of these components are constant.
According to the Doppler effect the wavelength of the radiation emitted in
the undulator on the first harmonic at the angle $\theta $ to the axis "y" is
              \begin{equation}  
              \lambda _1= {\lambda _u\over 2\gamma ^2}(1 + p_{\perp}^2 +
              \vartheta ^2),              \end{equation}
where $p_{\perp} = \gamma \beta _{\perp} = eB_m\lambda _u/2\pi mc^2 \simeq
B_m[Gs]\lambda _u [cm]/10700$ is the deflecting parameter of the\\
undulator\footnote[7]{Deflecting parameter
of the undulator is the product of rootmeansquare of the banding angle
$\alpha $ and $\gamma $. Approximately $p _{\perp} \simeq \alpha _m\gamma$,
where $\alpha _m$ is the maximum deflecting angle of the electron in the
undulator.}, $B_m$ is the value of the magnetic field of the undulator,
$\vartheta = \gamma \theta$.
           \vskip 10mm \hskip 32mm
\unitlength=0.40mm
\linethickness{0.4pt}
\begin{picture}(258.00,110.00)
\put(-70.00,50.00){\line(1,0){320.00}}
\put(250.00,50.00){\vector(1,0){8.00}}
\put(0.00,10.00){\line(0,1){80.00}}
\put(200.00,40.00){\line(0,1){20.00}}
\put(200.00,60.00){\line(-1,2){9.00}}
\put(200.00,40.00){\line(-1,-2){9.00}}
\put(191.00,78.00){\line(-1,1){30.00}}
\put(191.00,22.00){\line(-1,-1){30.00}}
\put(70.00,0.00){\line(0,1){100.00}}
\put(80.00,0.00){\line(0,1){100.00}}
\put(10.00,0.00){\line(0,1){100.00}}
\put(140.00,0.00){\line(0,1){100.00}}
\put(10.00,0.00){\line(1,0){60.00}}
\put(80.00,0.00){\line(1,0){60.00}}
\put(80.00,100.00){\line(1,0){60.00}}
\put(10.00,100.00){\line(1,0){60.00}}
\put(250.00,43.00){\makebox(0,0)[cc]{$y$}}
\put(-50.00,88.33){\makebox(0,0)[cc]{$e^{-}$}}
\put(217.50,20.00){\makebox(0,0)[cc]{$e^{-}$}}
\put(65.00,35.00){\makebox(0,0)[cc]{$e^{-}$}}
\put(45.00,107.00){\makebox(0,0)[cc]{\small BM1}}
\put(110.00,107.00){\makebox(0,0)[cc]{\small BM2}}
\put(45.00,7.00){\makebox(0,0)[cc]{\small $\vec B\oplus $}}
\put(110.00,7.00){\makebox(0,0)[cc]{\small $\vec B\odot $}}
\put(110.00,93.00){\makebox(0,0)[cc]{\small ME}}
\put(-10.00,20.00){\makebox(0,0)[cc]{\small M1}}
\put(200.00,90.00){\makebox(0,0)[cc]{\small M2}}
\put(0.00,15.00){\line(5,0){100.00}}
\put(100.00,15.00){\line(5,-1){40.00}}
\put(140.00,7.00){\line(2,-1){24.00}}
\put(0.00,85.00){\line(5,0){100.00}}
\put(100.00,85.00){\line(5,1){40.00}}
\put(140.00,93.00){\line(2,1){24.00}}
\put(140.00,54.29){\line(0,-1){0.95}}
\put(170.00,24.00){\makebox(0,0)[cc]{\small M3}}
\put(150.00,26.67){\line(1,-1){10.00}}
\put(154.17,18.33){\vector(-1,-4){10.83}}
\put(150.83,-27.67){\makebox(0,0)[lc]{\small Extracted radiation}}
\put(-50.00,-64.00){\makebox(0,0)[lc]{\small Fig.3: Schematic diagram of the
experimental setup. M1: plain mirror; M2: spherical}}
\put(-23.00,-80.00){\makebox(0,0)[lc]{\small mirror; M3: extracting mirror;
BMs: banding magnets; ME: mode envelope.}}
\put(10.00,50.00){\line(3,-1){15.33}}
\put(25.33,45.00){\line(5,-1){10.33}}
\put(48.67,43.00){\line(6,1){15.00}}
\put(82.33,54.70){\line(3,1){11.00}}
\put(93.33,58.30){\line(6,1){9.33}}
\put(110.33,60.00){\line(5,-1){9.67}}
\put(120.00,58.00){\line(4,-1){10.67}}
\put(131.00,55.00){\line(2,-1){9.67}}
\put(82.33,55.00){\line(-2,-1){18.33}}
\put(61.67,41.67){\vector(2,1){12.50}}
\put(102.67,59.67){\line(1,0){8.00}}
\put(48.67,42.67){\line(-1,0){13.00}}
\put(140.00,50.00){\line(2,-1){90.83}}
\put(221.67,16.67){\vector(2,-1){13.33}}
\put(10.00,50.00){\line(-2,1){80.00}}
\put(-70.00,90.00){\line(0,0){0.00}}
\put(-46.67,85.00){\vector(2,-1){12.50}}
\end{picture}
                 \vskip 40mm
Let short electron bunches enter the resonator in the time interval 
$T_b = T_w/m$, where $T_w = 2d/c$ is the period of the stored wave 
packet oscillations between the mirrors $M1$ and $M2$, $m = 1,2,3...$ 
is integer \footnote[8]{The scheme can work both on harmonic and on 
subharmonic of the bunching frequency or in general case under 
condition $nT_b = mT_w$, where $m,n$ are integer.}. In this case the 
train of the electron bunches will excite many longitudinal modes in 
the resonator such a way that the total wavepacket of the length 
$K\lambda _1$ will be installed after the time interval $QT_w$, where 
$\lambda _1$ is the wavelength of the emitted undulator radiation, $Q$ 
is the quality of the resonator, $K$ is the number of the undulator 
periods. Each next emitted wave packet will overlap the previous ones.

The quality of the resonator $Q$ is defined by the condition $Q^{-1} =
T_w(d\ln {\varepsilon ^{em}}/dt)$, where $\varepsilon ^{em} = (c/4\pi )\int
|\vec E|^2dSdt$ is the energy of the electromagnetic bunch stored
in the resonator at equilibrium, $d\varepsilon ^{em}/dt$ is the rate of the
electromagnetic energy loss in the resonator, $dS = 2\pi rdr$ is the
element of the area of the stored wavepacket. Approximately $Q^{-1} \simeq
Q^{-1}_{abs} + Q^{-1} _{diff}$, where $Q_{abs} \simeq 1/(1 - R_{M1}R_{M2})$
is the value defined by absorption in mirrors, $R_{M1}, R_{M2}$ are the
reflection coefficients of the mirrors, $Q_{diff}$ is the quality of the
resonator defined by diffraction losses of the resonator. Optimum regime of
the resonator excitation and extraction of the stored radiation corresponds to
the case $Q_{abs} = Q _{diff}, Q = Q_{abs}/2$ when diffraction losses are
defined mainly by the radiation extracted through the resonator hole or
through the mirror $M3$.

At equilibrium the energy losses per one period and per one bunch
of the radiation stored in the resonator $T_wd\varepsilon ^{em}/dt =
\varepsilon ^{em}/Q$ is equal to the energy $\Delta \varepsilon _b$ which is
transformed by the electron bunch to the wavepacket. In the case of the
point-like electron bunch of zero emittance, the value $\Delta \varepsilon _b
= eN_b\int \vec E \vec vdt$  where $N_b$ is the number of the electrons in a
bunch. At resonance the value $\int \vec
E\vec v dt = E_{\perp }v_{\perp }K\lambda _u/c = E_{\perp }\beta _{\perp }K
\lambda _u$, where symbols $\perp $ are related to the transverse components
of the corresponding vectors. In this case the electrons phases (angles
between $\vec E _{\perp}$ and $\vec v_{\perp}$) are zero (directions of the
electron velocity $\vec v_{\perp}$ and $\vec E_{\perp}$ coincide) and that is
why the maximum deceleration  of the electron bunches takes place.

For the pointlike electron bunches the vector of the electric field
strength of the electromagnetic bunch stored in the resonator can be presented
in the form
              \begin{equation} 
\vec E(t) = Re\{\vec e E_{\perp }f(t - t_w)e^{-r^2/\sigma _\gamma ^2}e^{-i
\omega _1(t - t_w)}\}
              \end{equation}
where $\vec e = \vec e_x - i\vec e_z$, vectors $\vec e_x, \vec e_z$ are the
unite vectors along axis $x,z$, $r^2 = x^2 + z^2$,
      $$E_{\perp } = {16\pi e N_bQ\gamma \over l_R \lambda _1}{p_{\perp}\over
       1 + p_{\perp}^2},               \hskip 10mm
              f(t - t_w) = \cases {1, &$0 \leq t - t_w \leq T$, \cr
       0, &$ t - t_w < 0, \; t - t_w > T,$} $$
$t$ is the current time, $t_w$ is the moment of the arrival of the wavepacket
to the coordinate $y$ under consideration, $T$ is the duration of the emitted
wave, Re $x$ is the real part of the number $x$. Usually the undulator has
the whole number of the periods $K$. In this case $T = KT_1$, where $T_1 =
\lambda _1(0)/c$ is the period of the wave emitted at the angle
$\theta = \vartheta = 0$.  The value $E_{\perp }$ can be defined from
the conditions of equilibrium:  $\varepsilon ^{em}/Q =
eN_bE_{\perp}\beta _{\perp}K\lambda _u$.

The energy $\Delta \varepsilon ^{out} = \varepsilon ^{em}/Q$ and the number
of the photons $\Delta N_{\gamma}^{out} = \Delta \varepsilon ^{out}/\hbar
\omega$ per one extracted wavepacket are equal
     \begin{equation}   
    \Delta \varepsilon ^{out} = {16\pi e^2QN_b^2\over \lambda _1}
   {K\lambda _u\over l_R}{p_{\perp}^2\over 1 + p_{\perp}^2}, \hskip 10mm
       \Delta N_{\gamma}^{out} = 8\alpha Q N_b^2{K\lambda
       _u\over l_R}{p_{\perp }^2\over 1 + p_{\perp}^2},
       \end{equation}
where $\alpha = e^2/\hbar c \simeq 1/137$ is the fine structure constant.

The pulsed power $P^{out} = \Delta \varepsilon ^{out}/T_b$ extracted from the
resonator according to (14) can be presented in the form
              \begin{equation}  
       P^{out} = 8\pi me^2c{QN_b^2K\lambda _u\over \lambda _1l_R d}
       {p_{\perp}^2\over 1 + p_{\perp}^2}
              \end{equation}
or $P^{out}[W] = 8.64\cdot 10^{-15}mQN_b^2(K\lambda _u/d)\lambda _1
^{-1}[cm]l^{-1}_R[cm]\,p_{\perp}^2/(1 + p_{\perp }^2)$.

It follows from (14),(15) that the number of the emitted photons and the total
power are proportional to the quality of the resonator and undulator length
but not to the number of the undulator periods. The value $K\lambda _u < d$
and $l_R \simeq d$. It means that for the case of the long wavelength broadband
sources of the electromagnetic radiation based on the relativistic electron
beams (when we are forced to choose $p_{\perp} \gg 1$ even in the case of
$\lambda _u \simeq d$) there is no necessity to increase the number of the
undulator periods by shorting period and increasing the deflecting parameter.
In this case we ought to choose $K \simeq 1\div 2, K\lambda _u \simeq d$.

When the electron bunches have the length comparable or large then the emitted
wavelength or when the bunches have high energy and angular spreads (large
emittances) then the coherence factor $F^{coh}$ and the square of the
coherence factor will appear in the equations (13) - (15)
(see Appendix 1). This factor can be derived in the following way. According
to Parseval's theorem $\int _{-\infty}^{\infty }f^2(t)dt = \int
_0^{\infty }|f_{\omega }|^2d\omega/\pi$. That is why the energy of the
wavepacket in the resonator can be presented in the form $\varepsilon
^{em} = (c/4\pi ^2)\int _0^{\infty}|\vec E_{\omega }|^2dSd\omega $. For
the point-like bunch the Fourrier's component $\vec E_{\omega }$ is the
known value. We have calculated the value $\varepsilon ^{em}$ through
the vector $\vec E(t)$ and determined the necessary parameters of the
emitted radiation (13)-(15) for this case.  In the case of the extended
electron beam the value $\vec E_{\omega} = \vec E_{\omega
1}i_{\omega}/e$, where $i_{\omega}$ is the Fourrie's transform of the
bunch current (see Appendix 1) and $\vec E_{\omega 1}$ is the Fourrie's
transform of the electric field strength of the radiation emitted by one
arbitrary electron of the beam \cite{bes3}. That is why in this case
the emitted power will be determined by the expression
              \begin{equation}  
       P^{out} = 8\pi me^2c{QN_b^2(F^{coh})^2K\lambda _u\over \lambda _1l_R d}
       {p_{\perp}^2\over 1 + p_{\perp}^2}
              \end{equation}
where $(F^{coh})^2 = (2\pi /e)^2\int _0^{\infty}|\vec E_{\omega 1}i_{\omega
}|^2d\omega /N_b^2\int |\vec E_{\omega 1}|^2d\omega $ (the vector $\vec
E_{\omega 1}$ can be substituted by a more simple vector $\vec \psi _{\omega
1}$ (see Appendix 1)). In the case of the point-like electron bunches the
values $2\pi i_{\omega }/e = N_b$, $F^{coh} = F^{coh}_{max} = 1$.

\vskip 6mm

$\underline{Example}$. Let be $Q = 10$, $N_b = 10^7$, $m = 14$, $K =
1$, $\lambda _u = 50$ cm, $l_R = d = 70$ cm, $\gamma = 80$, $\lambda = 1$ mm,
$F^{coh} = 1$. In this case according to (15): $p_{\perp} \simeq 5$, $B_m =
1070$ Gs, $P^{out} = 25$ W.
                  \vskip 6mm
\section{Transition radiation sources.}

\subsection{Introduction to the theory of the transition radiation.}

The transition radiation appear when an electron or other charged particle
pass through (cross) a boundary between two media with different indexes of
refraction. The shape and dimensions of the boundary can be arbitrary. A
plane metal mirror in vacuum is one of the examples of such boundary
(see Fig.4). The transition radiation in the real mirrors is emitted from a
small ($\sim 10^{-5}$ cm) regions in the form of semi-sphere layers of the
same thickness originated from the points of the electron entrance in and
exit from the mirror.

The image method of calculation of the transition radiation can be used when
the mirror have infinite dimensions and conductivity. In this case the
transition radiation of the falling electron is equivalent to the strange
radiation emitted by the electron and positron which moved from infinity in
the opposite directions with equal velocities and stopped at the boundary of
the mirror. The transition radiation of the electron emerging from the mirror
corresponds to the instantaneous start of the electron and it's image at the
mirror surface \footnote[9]{Notice that both the angular
dependence and the direction (sign) of the electric field strength of the
emitted wave are the same for the cases of the electron start and finish at
the mirror surface.}. In this case the spectral-angular distribution of the
emitted radiation will be determined by the eq-s (4),(6) for the case
$\beta _1 = 0$ and for opposite signs of the charges and directions of
the electron and it's image velocities
              \begin{equation}  
              {\partial ^2\varepsilon \over \partial {\omega }\partial o} =
              {e^2\beta ^2\over \pi ^2 c}{\sin ^2\theta \over (1 -
              \beta ^2\cos ^2\theta)^2}|_{\theta \ll 1} = {e^2\gamma ^2\over
              \pi ^2 c}{\vartheta ^2\over (1 + \vartheta ^2)^2},
              \end{equation}
where $0 \leq \theta \leq \pi /2$ is the angle between the direction of the
electron velocity and the direction from the electron's start point to the
observation point.

The spectral-angular distribution of the emitted radiation gathered over the
azimuth angle is
              \begin{equation}  
              {\partial ^2\varepsilon \over \partial {\omega }\partial \theta
              } = {2e^2\beta ^2\over \pi c}{\sin ^3\theta \over (1 -
              \beta ^2\cos ^2\theta)^2}|_{\theta \ll 1} \simeq {2e^2\beta
              ^2\over \pi c}{\gamma \vartheta ^3\over (1 + \vartheta ^2)^2}
                          \end{equation}
where $\vartheta = \gamma \theta $. It follows from (18) that the emitted
radiation is distributed in a broad range of angles up to the angle $\theta
= \pi /2$. A broad peak exists at the angle $\theta = \sqrt 3/\gamma$.

The spectral distribution of the radiation emitted over the range of the
angles $\theta _1$, $\theta _2$ is
              \begin{equation}  
              {\partial \varepsilon \over \partial {\omega }}(\theta _1,
              \theta _2) = {e^2\over \pi c}[{1 + \beta ^2\over 2\beta}\ln
              {(1 - \beta \cos \theta _2)(1 + \beta \cos \theta _1)
              \over (1 + \beta \cos \theta _2)(1 - \beta \cos \theta _1)} -
              {1\over \gamma ^2}({\cos \theta _1 \over 1 - \beta ^2\cos ^2
           \theta _1} - {\cos \theta _2 \over 1 - \beta ^2 \cos ^2\theta _2})].
              \end{equation}
In particular the spectral distribution of the transition radiation emitted
in the range of the angles $\theta _1 \leq \theta \leq \pi /2$ and the
spectral distribution of the total radiation emitted in the range of angles
$0 \leq \theta \leq \pi /2$ are\\

              \begin{equation}  
              {\partial \varepsilon \over \partial {\omega }}(\theta _1,
              \theta _2)|_{\theta _2 = \pi /2}  = {e^2\over \pi c}[{1 +
              \beta ^2\over 2\beta}\ln {(1 + \beta \cos \theta _1)
              \over (1 - \beta \cos \theta _1)} - {1\over \gamma ^2}({\cos
              \theta _1 \over 1 - \beta ^2\cos ^2 \theta _1})],
              \end{equation}

              \begin{equation}  
              {\partial \varepsilon \over \partial {\omega }} =
              {e^2\over \pi c}[{1 + \beta ^2\over \beta }
              \ln (1 + \beta)\gamma - 1]|_{\gamma \gg 1} \simeq {2e^2\over
              \pi c}[\ln 2\gamma - 1].              \end{equation}

\begin{center}\vskip 15mm
\special{em:linewidth 0.4pt}
\unitlength 1.00mm
\linethickness{0.4pt}
\begin{picture}(76.00,111.00)
\put(40.00,44.67){\line(0,1){60.00}}
\put(40.00,74.67){\circle*{0.70}}
\put(40.00,99.78){\line(6,-1){7.33}}
\put(47.33,98.67){\line(5,-2){6.00}}
\put(53.33,96.22){\line(6,-5){5.78}}
\put(59.11,91.33){\line(2,-3){3.78}}
\put(62.89,85.56){\line(1,-3){2.44}}
\put(65.33,78.44){\line(0,-1){6.22}}
\put(65.33,72.22){\line(-1,-4){1.78}}
\put(63.56,64.67){\line(-2,-3){3.78}}
\put(59.78,59.11){\line(-1,-1){4.67}}
\put(55.11,54.44){\line(-5,-3){4.89}}
\put(50.22,51.56){\line(-3,-1){5.56}}
\put(44.67,49.78){\line(-1,0){8.44}}
\put(36.22,49.78){\line(-4,1){6.44}}
\put(29.78,51.33){\line(-3,2){4.00}}
\put(25.78,54.00){\line(-6,5){3.56}}
\put(22.22,56.89){\line(-5,6){3.33}}
\put(18.89,60.89){\line(-1,2){2.44}}
\put(16.44,65.78){\line(-1,4){1.33}}
\put(15.11,71.56){\line(0,1){6.00}}
\put(15.11,77.56){\line(1,5){1.11}}
\put(16.22,82.89){\line(1,2){2.00}}
\put(18.22,87.11){\line(2,3){2.44}}
\put(20.67,90.89){\line(1,1){4.44}}
\put(25.11,95.33){\line(2,1){4.22}}
\put(29.33,97.33){\line(3,1){5.78}}
\put(35.11,99.33){\line(1,0){4.89}}
\put(40.00,41.00){\makebox(0,0)[cc]{M}}
\put(39.33,30.67){\makebox(0,0)[cc]
{Fig.4:\small The scheme of the transition  radiation of the electron
passing through a plain mirror; }} \put(43.50,26.00){\makebox(0,0)[cc]
{\small $\vec E^c$is the Coulomb electric  field strength lines of the
electron; arrows near vectors $\vec E^r$ show the}}
\put(43.33,21.00){\makebox(0,0)[cc]
{\small directions of the electric field strength lines of the radiation
propagating in the spherical layer.}} \put(59.33,74.67){\circle{0.67}}
\put(62.66,76.97){\makebox(0,0)[cc]{$e^{-}$}}
\put(59.33,74.67){\line(0,1){8.33}}
\put(59.33,74.67){\line(0,1){0.00}}
\put(59.33,74.67){\line(1,6){1.33}}
\put(59.33,74.67){\line(-1,5){1.67}}
\put(57.67,82.67){\line(-2,5){2.33}}
\put(55.33,88.33){\line(-5,6){3.00}}
\put(52.33,92.00){\line(-5,3){4.67}}
\put(47.67,94.67){\line(-4,1){5.33}}
\put(65.33,74.67){\vector(-1,0){6.00}}
\put(40.00,74.67){\vector(1,0){19.00}}
\put(59.33,74.67){\line(-3,4){4.00}}
\put(55.33,80.00){\line(-3,2){3.67}}
\put(51.67,82.33){\line(-3,1){6.00}}
\put(45.67,84.33){\line(-1,0){2.67}}
\put(59.33,74.67){\line(-3,-4){3.33}}
\put(56.00,70.33){\line(-3,-2){5.00}}
\put(51.00,67.00){\line(-4,-1){4.67}}
\put(59.00,74.67){\line(-1,-4){2.67}}
\put(56.33,64.67){\line(-3,-5){4.33}}
\put(52.00,57.67){\line(-5,-4){4.67}}
\put(59.33,74.67){\line(0,-1){9.33}}
\put(59.33,74.67){\line(1,-6){1.67}}
\put(47.33,54.00){\line(-3,-1){5.67}}
\put(59.33,83.00){\line(-1,6){1.20}}
\put(45.67,88.67){\makebox(0,0)[cc]{\small $\vec E^c$}}
\put(40.00,100.33){\line(6,-1){8.33}}
\put(48.33,99.00){\line(5,-2){6.00}}
\put(54.33,96.67){\line(1,-1){5.67}}
\put(60.00,91.00){\line(3,-4){3.33}}
\put(63.33,86.33){\line(1,-3){2.00}}
\put(65.33,80.67){\line(1,-5){1.33}}
\put(66.67,73.67){\line(-1,-4){2.00}}
\put(64.67,65.00){\line(-3,-5){4.00}}
\put(60.67,58.33){\line(-5,-4){9.00}}
\put(51.67,51.00){\line(-3,-1){6.67}}
\put(45.00,48.67){\line(-1,0){10.00}}
\put(35.00,48.67){\line(-5,2){9.33}}
\put(25.67,52.33){\line(-1,1){5.33}}
\put(20.33,57.67){\line(-2,3){5.00}}
\put(15.33,65.33){\line(-1,5){1.33}}
\put(14.00,71.33){\line(0,1){7.67}}
\put(14.00,79.00){\line(2,5){4.00}}
\put(18.00,88.67){\line(5,6){5.33}}
\put(23.33,95.00){\line(2,1){10.67}}
\put(34.00,100.33){\line(1,0){6.00}}
\put(9.33,108.67){\line(1,0){66.33}}
\put(43.00,111.00){\makebox(0,0)[cc]
{\small Spherical layer of the transition radiation}}
\put(9.33,108.67){\vector(1,-2){10.00}}
\put(76.00,108.67){\vector(-1,-2){12.00}}
\put(14.00,62.67){\vector(-1,3){1.67}}
\put(14.00,87.00){\vector(-1,-3){1.67}}
\put(66.67,87.00){\vector(1,-3){1.33}}
\put(66.33,62.67){\vector(1,3){1.67}}
\put(69.67,63.33){\makebox(0,0)[cc]{$\vec E^r$}}
\put(71.67,85.33){\makebox(0,0)[cc]{$\vec E^r$}}
\put(10.33,85.33){\makebox(0,0)[cc]{$\vec E^r$}}
\put(10.33,63.00){\makebox(0,0)[cc]{$\vec E^r$}}
\put(40.00,96.00){\line(1,0){2.33}}
\put(43.00,84.33){\line(-1,0){3.00}}
\put(40.00,65.67){\line(1,0){6.33}}
\put(42.00,52.00){\line(-1,0){2.00}}
\put(61.00,64.67){\line(0,-1){4.00}}
\put(59.33,65.00){\line(-1,-5){1.67}}
\put(60.67,82.67){\line(0,1){6.00}}
\put(58.13,89.00){\line(-1,4){1.10}}
\put(53.00,59.00){\vector(2,3){0.67}}
\put(52.00,67.67){\vector(3,2){1.00}}
\put(53.00,81.33){\vector(3,-2){1.00}}
\put(53.33,91.00){\vector(2,-3){0.67}}
\put(58.13,89.00){\vector(1,-4){0.13}}
\put(58.67,61.00){\vector(1,4){0.33}}
\put(61.00,64.33){\vector(0,1){0.33}}
\put(60.67,83.67){\vector(0,-1){0.67}}
\emline{22.00}{94.00}{1}{23.00}{102.00}{2}
\put(25.00,108.00){\vector(1,3){0.2}}
\emline{23.00}{102.00}{3}{25.00}{108.00}{4}
\emline{26.00}{97.00}{5}{28.00}{103.00}{6}
\put(31.00,108.00){\vector(2,3){0.2}}
\emline{28.00}{103.00}{7}{31.00}{108.00}{8}
\emline{33.00}{100.00}{9}{35.00}{105.00}{10}
\put(38.00,108.00){\vector(1,1){0.2}}
\emline{35.00}{105.00}{11}{38.00}{108.00}{12}
\emline{33.00}{49.00}{13}{35.00}{44.00}{14}
\put(38.00,41.00){\vector(1,-1){0.2}}
\emline{35.00}{44.00}{15}{38.00}{41.00}{16}
\emline{26.00}{41.00}{17}{26.00}{52.00}{18}
\emline{26.00}{52.00}{19}{28.00}{45.00}{20}
\put(31.00,41.00){\vector(3,-4){0.2}}
\emline{28.00}{45.00}{21}{31.00}{41.00}{22}
\emline{22.00}{56.00}{23}{23.00}{47.00}{24}
\put(25.00,43.00){\vector(1,-2){0.2}}
\emline{23.00}{47.00}{25}{25.00}{43.00}{26}
\emline{17.00}{63.00}{27}{14.00}{49.00}{28}
\put(15.00,44.00){\vector(1,-4){0.2}}
\emline{14.00}{48.00}{29}{15.00}{44.00}{30}
\emline{14.00}{49.00}{31}{14.00}{48.00}{32}
\emline{17.00}{86.00}{33}{14.00}{101.00}{34}
\put(15.00,105.00){\vector(1,4){0.2}}
\emline{14.00}{101.00}{35}{15.00}{105.00}{36}
\put(3.00,75.00){\vector(-1,0){0.2}}
\emline{14.00}{75.00}{37}{3.00}{75.00}{38}
\put(19.00,102.00){\makebox(0,0)[cc]{$\vec E^c$}}
\put(19.00,48.00){\makebox(0,0)[cc]{$\vec E^c$}}
\put(7.00,71.00){\makebox(0,0)[cc]{$\vec E^c$}}
\end{picture}
\vskip -5mm\end{center}
It follows from the equations (19)-(21) that total radiation increases slowly
with the relativistic factor $\gamma $ ($\sim \ln \gamma$). A marked part of
the emitted radiation is emitted outside of the peak $\theta > \sqrt 3/\gamma
$. The energy of the transition radiation in the region $\gamma \gg 1/\theta
_1$ does not depend on $\gamma $.

Uniformly moving particle does not emit radiation. But such particle
in ultrarelativistic case produce electric and magnetic fields, which
have mainly transverse components of electric and magnetic fields (lines
of force are compressed in the transverse direction). This fields obviously
will interact with matter the same way as the real wavepackets of the
electromagnetic waves of the same time dependence of the electric
and magnetic fields at the observation point. One speaks that the Coulomb
fields of particles present the virtual photons (quanta). Virtual photons can
be considered as real photons moving parallel to the electron velocity.
They can excite or ionize atoms and molecules or can be reflected from the
mirror when electron pass by the mirror.

If the relativistic electron hit the
mirror of finite dimensions and is stopped in it then in the first
approximation the virtual photons which are outside the mirror dimensions
will be emitted in the direction of the initial electron velocity and the
other virtual photons will be reflected from the mirror in the opposite
direction.  The properties of the virtual photons are presented in the
Appendix 2.

If the electron pass through the mirror of finite dimensions far from the edge
of the mirror then the interference of the transition radiation emitted from
the both sides of the mirror and diffraction radiation will occur at the
observation point. This interference can be essential in the long wavelength
region. The total energy of the transition radiation weekly depend on the
form and dimensions of the mirror and depends mainly on the energy of the
emerged spherical layers. The form of the mirror have influence on the
distribution of the transition radiation. For example, the transition
radiation will be focused by the concave mirror \cite{riazanov}.

Notice that the expression (17) was calculated by using the image method. In
this case the problem was solved exactly. At the same time we can consider
this problem in the framework of virtual photons \cite{jackson}. In this case
we can say that the relativistic electron brings the compressed in the
longitudinal direction electromagnetic field in the form of a wavepacket
and this field will be reflected from the mirror after the electron will
disappear in it. This is the approximate method \cite{jackson}. That is why
the expressions (21) and (37) are slightly differ. Moreover the transition
radiation originate from one point of the mirror surface and propagate in
the form of thin spherical layer with sharp boundaries (in the case of the
mirror with the infinite conductivity the thickness of the layer will be
zero) as we discussed above using exact image method. The solution obtained
by the approximate method of the virtual photons in any case lead to the
plane (not spherical) reflected wavepackets of finite thickness which at the
mirror surface have the electric field strengths with time dependence of the
form which repeat the form of the Coulomb fields of the electron.

As we mentioned the transition radiation problem is equivalent to the
instantaneous finish at (instantaneous start from) the point of input (output)
of the particle and its image on the mirror surface. For the ideal mirror the
emitted radiation is the strange radiation and for the real mirror is the
conventionally strange radiation (after strange wavepacket a week but long
wavepacket of opposite polarity with the same strange parameter appear)
because in the last case the waves will be emitted by electrons of the mirror
but this electrons will be at rest both before and after transition of
the projectile particle the mirror (real mirrors have finite conductivity)
\cite{bes1a}, \cite{bes1b}, \cite{bes3}. Notice that the finite dimensions
and conductivity of the mirror will lead to decreasing of the long wavelength
and short wavelength radiation accordingly.

          \subsection{The transition radiation of particles crossing
          a series of two developed at the angle 45$^o$ mirrors.}

In the Fig.5 a scheme of the transition radiation source based on the system
of two developed at the angle 45$^o$ mirrors is presented \cite{x},
\cite{shib1}. The particle come out of the mirror $M1$ at the point $A$ with
the velocity
$v$ moves along the axis $y$ and strike the mirror $M2$ at the point $O$. At
the point of observation $D$ the radiation will be received directly both
from the point $A$ and from the point $O$. There are many other ways for the
propagation of the radiation emitted in the points $A$ and $O$ which lead to
the observation point $D$ after many reflections between mirrors $M1$ and
$M2$. Further we will consider the case when the distances $R_{AD}$,
$R_{OD}$ between the observation point $D$ and mirror's points $A$, $O$
is of the order of the distance $R_{AO}$ between points $A$ and $O$. In
this case we will neglect the radiation emitted in the point $A$ and
propagated to the point $D$ under the sharp angle. We will neglect the
radiation emitted in the points $A$ and $O$ and reflected by mirrors
$M1$ and $M2$ when this radiation propagate to the point $D$ at sharp
angles as well. We will select only the radiation which propagate to
the observation point $D$ along the axis $x$ and at small angles to
this axis.

Using the image method we can say that at the observation point
$D$ the next wavepackets of such kind will be received. The first wavepacket
will be presented by the radiation emitted from the point $A^{'}$ by the
image $-e$ and the image of the image $e$ (this is the transition radiation
emitted in the point $A$ and reflected by the mirror $M2$). The second one
will be presented by the radiation emitted from the point $O$ by the stopped
charge $e$ moved in the $y$ direction and its image $-e$ moved in the
direction $x$ (this is the transition radiation emitted in the point $O$).
The third one will be presented by the radiation emitted from the point
$O^{''}$ placed on the axis $x$ at the distance $OO^{''} = 2OA^{'}$ where the
images of the images will be stopped when the charge will strike the mirror
$M2$ (this is the transition radiation emitted by charge in the point $O$ in
the direction opposite to it's velocity and reflected successively by the
mirrors $M1$ and $M2$).
     \vskip 10mm\begin{center}
\unitlength=0.40mm
\linethickness{0.4pt}
\begin{picture}(262.38,309.67)
\put(76.19,160.15){\line(1,0){183.81}}
\put(100.48,184.73){\line(0,-1){49.77}}
\put(119.00,160.22){\line(1,0){9.67}}
\put(119.67,160.22){\line(1,0){2.33}}
\put(149.05,184.12){\line(1,-1){50.48}}
\put(169.05,77.20){\line(0,-1){13.52}}
\put(169.05,63.69){\line(1,0){8.57}}
\put(177.62,63.69){\line(0,1){12.90}}
\put(177.62,76.59){\line(-1,0){8.57}}
\put(172.86,228.42){\line(0,1){6.38}}
\put(172.86,236.34){\line(0,1){5.38}}
\put(172.86,239.80){\line(0,1){5.68}}
\put(172.86,227.13){\line(0,-1){6.14}}
\put(172.86,219.75){\line(0,-1){6.14}}
\put(172.86,212.38){\line(0,-1){6.14}}
\put(172.86,205.62){\line(0,-1){6.14}}
\put(172.86,198.25){\line(0,-1){6.14}}
\put(172.86,190.88){\line(0,-1){6.14}}
\put(172.86,183.50){\line(0,-1){6.14}}
\put(172.86,176.13){\line(0,-1){6.14}}
\put(172.86,168.14){\line(0,-1){6.14}}
\put(172.86,298.00){\line(0,1){5.68}}
\put(172.86,296.96){\line(0,-1){5.14}}
\put(172.86,290.54){\line(0,-1){5.14}}
\put(172.86,285.38){\line(0,-1){5.14}}
\put(172.86,279.35){\line(0,-1){5.14}}
\put(172.86,273.33){\line(0,-1){5.14}}
\put(172.86,267.31){\line(0,-1){5.14}}
\put(172.86,261.29){\line(0,-1){8.14}}
\put(172.86,252.29){\line(0,-1){8.14}}
\put(172.86,160.15){\vector(0,-1){78.65}}
\put(172.86,160.15){\circle*{2.46}}
\put(123.33,169.37){\vector(1,0){14.76}}
\put(100.95,159.54){\circle*{2.46}}
\put(94.76,153.39){\makebox(0,0)[cc]{$A$}}
\put(179.52,166.91){\makebox(0,0)[cc]{$O$}}
\put(260.00,152.78){\makebox(0,0)[cc]{$y$}}
\put(260.00,160.15){\vector(1,0){2.38}}
\put(173.33,69.83){\makebox(0,0)[cc]{$D$}}
\put(100.00,125.75){\makebox(0,0)[cc]{$M1$}}
\put(200.00,125.75){\makebox(0,0)[cc]{$M2$}}
\put(103.81,157.08){\vector(3,-4){54.76}}
\put(157.14,232.04){\makebox(0,0)[cc]{$A'$}}
\put(130.00,173.12){\makebox(0,0)[cc]{$e, \vec v$}}
\put(119.50,160.22){\oval(1.67,4.30)[]}
\put(181.00,86.00){\makebox(0,0)[cc]{$x$}}
\put(172.86,302.55){\circle*{2.46}}
\put(172.86,234.62){\circle*{2.46}}
\put(156.67,302.58){\makebox(0,0)[cc]{$A^{''}$}}
\put(172.78,249.96){\line(0,1){5.02}}
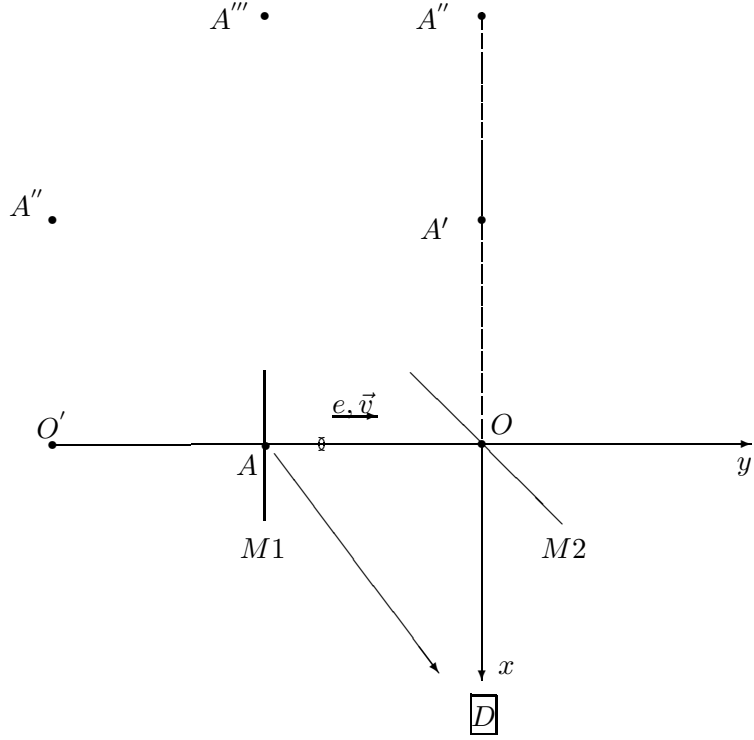
\put(110.00,22.00){\makebox(0,0)[cc]{\small Fig.5: A transition radiation
scheme based on two turned at the angle 45$^o$ mirrors. Electron}}
\put(114.00,12.00){\makebox(0,0)[cc]{\small goes along the axis $y$ and pass
through  mirrors $M1$ and $M2$ at points $A$ and $O$.  Points}}
\put(114.00,2.00){\makebox(0,0)[cc]{\small $A^{'}$, $A^{''}$,
$A^{'''}$ ... and $O^{'}$, $O^{''}$... are the images and images of images
of the points $A$ and $O$.}}
\put(76.00,160.00){\line(-1,0){46.00}}
\put(30.00,160.00){\circle*{2.46}}
\put(30.00,166.67){\makebox(0,0)[cc]{$O^{'}$}}
\put(30.00,234.67){\circle*{2.46}}
\put(21.67,240.67){\makebox(0,0)[cc]{$A^{''}$}}
\put(100.67,302.67){\circle*{2.46}}
\put(89.00,302.67){\makebox(0,0)[cc]{$A^{'''}$}}
\end{picture}
\end{center}\vskip 5mm
If we mark the unit vectors along the axis $y$ and $x$ through $\vec e_y$
and $\vec e_x$ accordingly then the velocity vectors of charge, it's image
and the image of the image will be $c\beta \vec e_y$, $c\beta \vec e_x$ and
$-c\beta \vec e_x$ respectively. Using this notations we can present the
Fourrier's component (5) in the form\\

                   \begin{equation}  
           \vec E_{\omega} = \sum _i\vec E_{\omega 1i}e^{i\omega t_{1i}},
                   \end{equation}
where subscriptions $i$ mean the emission points $A^{'}$, $O$, $O^{''}$ and
the moments $t_{1i}$ mean the moments when the radiation emitted at these
points come to the observation point $D$, $\vec E_{\omega 1i} = (1/2\pi )\vec
I_i$,\\

    $$\vec I_1 = {e\over cR_{A^{`}D}}[\vec n[\vec n(
    {\beta \vec e_x\over 1 - \beta \cos \theta _{A^{`} x}}
    + {\beta \vec e_x\over 1 + \beta \cos \theta _{A^{`} x}})]],$$

    $$\vec I_2 = -{e\over cR_{OD}}[\vec n[\vec n(
    {\beta \vec e_x\over 1 - \beta \cos \theta _{O x}}
    + {\beta \vec e_y\over 1 + \beta \cos \theta _{O y}})]], $$

    $$\vec I_3 = -{e\over cR_{O^{``}D}}[\vec n[\vec n(
    {\beta \vec e_x\over 1 + \beta \cos \theta _{O^{``} x}}
    - {\beta \vec e_x\over 1 - \beta \cos \theta _{O^{``} y}})]], $$
$R_{A^{'}D}, R_{OD}, R_{O^{''}D}$ are the distances between points $A^{'}, O,
O^{''}$ and the observation point, $\cos \theta _{A^{'} x}\\ = \vec n_A \vec
e_x$ $\cos \theta _{O x} = \vec n_O \vec e_x$, $\cos \theta _{O^{''} x} = \vec
n_O^{''} \vec e_x$, $\cos \theta _{Oy} = \vec n_O \vec e_y$, $\cos \theta
_{O^{''} y} = \vec n_{O^{''}} \vec e_y$, $\vec n_i$ are the unit vectors
directed from the point $i$ to the observation point.

It follows from (22) that at the distances between the emission points $A$ and
$O$ of the order of the distance between the point $O$ and the observation
point $D$ or higher ($R_{AO} \geq R_{OD}$) the density and spectrum of the
radiation accepted by the observer will be determined in the main by the
radiation emitted at the point $O$. In this geometry the long wavelength
radiation at the observation point will be presented in the highest degree.
The using of the concave mirror at this position will permit to select the
radiation in points $A$ and $O$ and to keep the long wavelength
radiation at high level. Such conditions probably took place in the
experiments \cite{shib1}.

At large distances $R_{OD} \gg R_{AO}$
the emitted radiation will be determined by the first terms of the vectors
$\vec I_1$ and $\vec I_2$ and the time difference between the arriving
moments to the observation point of the wavepackets of the transition
radiation emitted on the mirrors $M1$ and $M2$ at the points $A$ and $O$ will
be equal to $t_{12} - t_{11} = R_{AO}(1 - \beta \cos \theta)/\beta c \simeq
(R_{AO}/2\gamma ^2c)(1 + \vartheta ^2)$. In this case according to (4), (22)
the spectral-angular distribution of the transition radiation
             \begin{equation}  
              {\partial ^2\varepsilon \over \partial {\omega }\partial o} =
              {e^2\beta ^2\over \pi ^2 c}{\sin ^2\theta \sin ^2[\omega
              /\omega _c(\theta)]\over (1 - \beta \cos \theta)^2}|_{\gamma
              \gg 1,\, _{\theta \ll 1}} \simeq {4e^2\gamma ^2\over \pi ^2c}
              {\vartheta ^2\sin ^2[(\omega /\omega _{co})(1 + \vartheta ^2)]
              \over (1 + \vartheta^2)^2},
              \end{equation}
where $\omega _c(\theta) = 2\pi c \beta/R_{AO}(1 - \beta \cos \theta)|_{\gamma
\gg 1,\,_{\theta \ll 1}} \simeq \omega _{co}/(1 + \vartheta ^2)$, $\omega
_{co} = 4\pi c\gamma ^2/R_{AO}$.

The multiple $\sin ^2[\omega /\omega _c(\theta)]$ in (23) appeared as the
result of the interference of two strange waves with opposite strange
parameters (the wave of the transition radiation emitted at the point $A$
change the direction of the electric field strength after the reflection
from the mirror M2). The interference does not change the total energy of the
radiation emitted in the given direction (and over all directions as well)
since the integration  of (23) over $\omega $ from $0$ to $\omega \gg \omega
_{co}$ will lead to double value of (17) ($<\sin ^2[\omega /\omega _{c}
(\theta )]> = 1/2$. The interference only redistribute the emitted energy
over the spectrum. In this case and usually in many other cases when a system
of similar mirrors placed in series and developed at large angles is used such
redistribution lead to the exhaustion of the emitted radiation in the
long wavelength region and the problem arises to select the emitted
wavepackets by lenses, holes and so on in order to destroy the
interference of the wavepackets of the opposite polarity at large distances
\cite{bes1a}, \cite{bes1b}.

The spectral distribution of the transition radiation gathered over the
azimuth angle and in the range of angles $0 \leq \theta \leq \theta _2$ is
             \begin{equation}  
             {\partial ^2\varepsilon \over \partial {\omega }\partial \theta}
              = {4e^2\over \pi c}\int _{\omega /\omega _{co}}^{\omega (1 +
              \vartheta _2^2)/\omega _{co}}(x - {\omega \over \omega _{co}})
              {\sin ^2x\over x^2}dx,
              \end{equation}
where $x = (\omega /\omega _{co})(1 + \vartheta ^2) = R_{AO}(1 + \vartheta
^2)/2\lambda \gamma ^2.$

In the long wavelength region $\lambda \gg R_{AO}(1 + \vartheta ^2)/2\gamma
^2|_{\vartheta \gg 1} = R_{AO}\theta ^2/2  $ and at small angles $\theta \ll
1$ ($x \ll 1$)
             \begin{equation}  
              {\partial \varepsilon \over \partial \omega } =
              {2e^2\over \pi c}({\omega \over \omega _{co}})^2\vartheta _2^4.
              \end{equation}

It follows from (23)-(25) that at large distances to the observation point
the emitted radiation in the long wavelength region will be suppressed the
less the higher angles
$\vartheta $ of the emitted radiation. The efficiency of the long wavelength
schemes based on the extraction of the transition radiation in the
neighboring mirrors will be searched at the next
part of the paper like the schemes based on the separation of the wavepackets
of the transition radiation.


    \subsection{Transition radiation of particles passing through a system of
mirrors.}

In the Fig.6 a system of two parallel circular mirrors arranged along the axis
$z$ is presented.\\

\vskip 10mm
\begin{center}
\unitlength=1.00mm
\linethickness{0.4pt}
\begin{picture}(123.67,140.00)
\put(30.00,30.00){\vector(0,1){110.00}}
\put(30.00,80.00){\circle{0.50}}
\put(10.00,75.00){\line(1,0){40.00}}
\put(50.00,85.00){\line(-1,0){40.00}}
\put(34.67,140.00){\makebox(0,0)[cc]{$z$}}
\put(14.33,88.00){\makebox(0,0)[cc]{$M2$}}
\put(14.33,72.00){\makebox(0,0)[cc]{$M1$}}
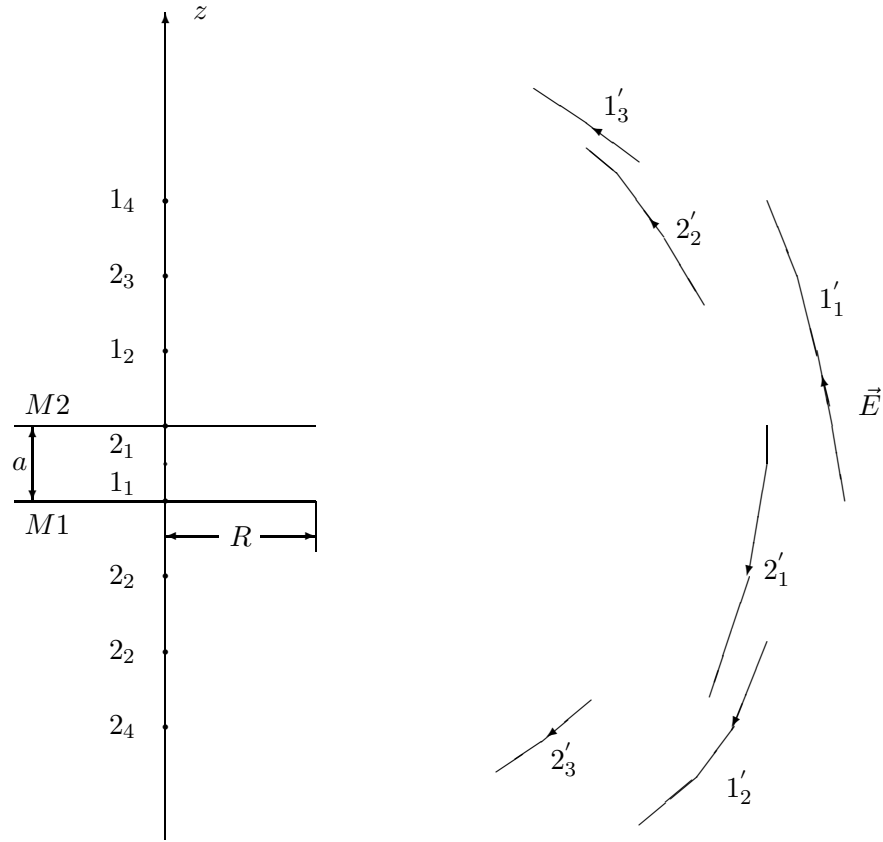
\put(60.33,20.00){\makebox(0,0)[cc]{\small Fig.6: The scheme of the transition
radiation of the electron passing through the system of two}}
\put(68.33,15.00){\makebox(0,0)[cc]{\small mirrors for the case of
$R = 2a$. Points $1_1$ and $2_1$ are the emission points of the transition}}
\put(68.33,10.00){\makebox(0,0)[cc]{\small radiation. Points $1_i$ and $2_i$
($i\ne 1$) are the images of the emission points. The figures $1^{'}_n$}}
\put(68.33,5.00){\makebox(0,0)[cc]{\small and $2^{'}_n$ (n = 1,2,3, ... )
mark the corresponding portions of the emitted wavepackets.}}
\put(79.00,130.00){\line(3,-2){7.00}}
\put(86.00,125.33){\line(4,-3){7.00}}
\put(86.00,122.00){\line(6,-5){4.00}}
\put(90.00,118.67){\line(3,-4){6.33}}
\put(96.33,110.00){\line(3,-5){5.33}}
\put(110.00,115.00){\line(2,-5){4.00}}
\put(114.00,105.00){\line(1,-4){2.67}}
\put(116.67,95.00){\line(1,-5){2.00}}
\put(118.67,85.00){\line(1,-6){1.67}}
\put(110.00,85.00){\line(0,-1){5.00}}
\put(110.00,80.00){\line(-1,-6){2.33}}
\put(107.67,65.00){\line(-1,-3){5.33}}
\put(110.00,56.33){\line(-2,-5){4.67}}
\put(93.00,32.00){\line(6,5){7.67}}
\put(100.67,38.33){\line(3,4){5.00}}
\put(74.00,39.00){\line(3,2){6.00}}
\put(80.00,43.00){\line(6,5){6.67}}
\put(106.33,37.00){\makebox(0,0)[cc]{$1_2^{'}$}}
\put(83.00,40.67){\makebox(0,0)[cc]{$2_3^{'}$}}
\put(111.33,66.00){\makebox(0,0)[cc]{$2_1^{'}$}}
\put(123.67,88.33){\makebox(0,0)[cc]{$\vec E$}}
\put(118.67,102.00){\makebox(0,0)[cc]{$1_1^{'}$}}
\put(99.67,111.33){\makebox(0,0)[cc]{$2_2^{'}$}}
\put(90.00,127.67){\makebox(0,0)[cc]{$1_3^{'}$}}
\put(10.67,80.00){\makebox(0,0)[cc]{$a$}}
\put(40.00,70.33){\makebox(0,0)[cc]{$R$}}
\put(42.67,70.33){\vector(1,0){7.33}}
\put(37.00,70.33){\vector(-1,0){7.00}}
\put(12.34,75.00){\vector(0,1){10.00}}
\put(12.34,75.67){\vector(0,-1){0.67}}
\put(50.00,75.00){\line(0,-1){6.67}}
\put(82.00,44.67){\vector(-4,-3){1.33}}
\put(106.00,46.33){\vector(-1,-2){0.67}}
\put(118.33,87.67){\vector(-1,4){1.00}}
\put(108.00,67.67){\vector(-1,-4){0.67}}
\put(95.33,111.67){\vector(-1,1){1.00}}
\put(88.00,124.00){\vector(-2,1){1.33}}
\put(24.33,115.00){\makebox(0,0)[cc]{$1_4$}}
\put(24.33,105.00){\makebox(0,0)[cc]{$2_3$}}
\put(24.33,95.00){\makebox(0,0)[cc]{$1_2$}}
\put(24.33,82.33){\makebox(0,0)[cc]{$2_1$}}
\put(24.33,77.33){\makebox(0,0)[cc]{$1_1$}}
\put(24.33,65.00){\makebox(0,0)[cc]{$2_2$}}
\put(24.33,55.00){\makebox(0,0)[cc]{$2_2$}}
\put(24.33,45.00){\makebox(0,0)[cc]{$2_4$}}
\put(30.00,45.00){\circle*{0.80}}
\put(30.00,55.00){\circle*{0.80}}
\put(30.00,65.00){\circle*{0.80}}
\put(30.00,95.00){\circle*{0.80}}
\put(30.00,105.00){\circle*{0.80}}
\put(30.00,115.00){\circle*{0.80}}
\put(30.00,115.00){\circle*{0.80}}
\put(30.00,75.00){\circle*{0.80}}
\put(30.00,85.00){\circle*{0.80}}
\end{picture}
\end{center}

\vskip 3mm
The electron pass trough mirrors in the direction of the axis $z$. In this
case two originating from points $1_1$ and $2_1$ thin spherical
wavepackets diverge between mirrors and go out by portions. First portion of
the spherical wavepacket emerged from the point $1$ go out in the upper
direction without reflection from the mirror $M2$ and propagates inside the
range of angles ($0, arctg \,(a/R)$), where $a$ is the gap between the
mirrors and $R$ is the radius of the mirrors. Second portion is reflected
from the mirror $M2$ and go out in the lower direction and propagate inside
the range of angles ($-arctg \,(a/R), -arctg \,(2a/R)$). The third portion
is reflected from the mirrors $M2$ and $M1$, go out in the upper direction
and propagates inside the range of angles ($arctg \,(2a/R), arctg \,(3a/R)$).
This process is continued to infinity. The directions of the electric field
strengths are changed after reflections from the mirrors. Propagation of the
radiation emitted in the point 2 can be considered by analogy.

We can see that at large distances the emitted wavepackets will not be
overlapped. The vector of the electric field strength of the radiation emitted
by electron will be directed backward to the hour-hand in the upper right part
of the Fig.6 and in the opposite direction in the lower right one. All emitted
transition radiation (twice of the value (21)) will be extracted from the
inner part of the system of mirrors (if we neglect the absorption of this
radiation in the
mirrors). The spectral-angular distribution of the emitted radiation can be
described by the expressions (17)-(21) because of the interference between the
extracted portions of radiation is absent. This conclusion will be valid for
the wavelengths $\lambda \ll 2a$.

In the Fig.7 a system of $M = 8$ circular parallel mirrors arranged along the
axis $z$ on equal distances and a conic mirror are presented. In this case all
gaps between mirrors will emit the identical wavepackets. If the distance
from the system to the observation point is higher then the length of the
system then the time dependence of the electric field strength $\vec E_n(t)$
of the wavepackets emitted from the gap "n" will differ from the first one
only by the time delay: $\vec E_n(t) = \vec E_1[t + (n-1)\Delta t]$, where
$\Delta t = \lambda _u(1 - \beta \cos \theta)/\beta c$, $\lambda _u$ is the
distance between mirrors, $\theta $ is the angle between the axis $z$ and the
direction to the observation point. The Fourrier's transforms of the vector
$\vec E_n(t)$ and the total vector $\vec E(t) = \sum _{n = 1}^{M}\vec E_n(t)$
will have the form $\vec E_{\omega n} = \vec E_{\omega _1}\exp [i\omega
(n-1)\Delta t]$ and $E_{\omega 1}[(\sin M\omega \Delta t/2)/(\sin \omega
\Delta t/2)]\exp [i\omega (M-1)\Delta t/2]$ respectively.

The spectral-angular distribution of the energy emitted in the system of
mirrors can be presented in the form
             \begin{equation}  
             {\partial ^2\varepsilon \over \partial {\omega }\partial o} =
             {\partial ^2\varepsilon _1\over \partial {\omega }\partial o}
              {\sin ^2(\pi M\omega /\omega _1)\over \sin ^2(\pi \omega
              /\omega _1},          \end{equation}
where ${\partial ^2\varepsilon _1/\partial {\omega }\partial o}$ is the
spectral-angular distribution of the energy emitted by one gap (consider
with (17)), $\omega _1 = 2\pi /\Delta t = 2\pi \beta c/\lambda _u(1 - \beta
\cos \theta)$.

It follows from (26) that at large distances the spectral-angular
distribution of the transition radiation emitted from the system of $2M$
mirrors have a line spectrum on the frequencies
             \begin{equation}  
              \omega _k = {k\omega _{1}\over 1 - \beta \cos \theta }.
              \end{equation}
where $k = 1,2,3 ... $ is the integer.
The width of the lines $\Delta \omega /\omega _k \simeq 1/kM$. The envelope
of the line spectrum is determined by the spectral-angular distribution of
the transition radiation (17) emitted from the surface of one mirror. At
given direction the spectral-angular and spectral distributions of the
emitted energy are increased with the number of mirrors by the low $M^2$ and
$M$ respectively.

Notice that the energy produced by a particle in the system of mirrors is a
maximum energy permitted for the transition radiation. It corresponds to the
case when the particle fall on the every mirror of the system as if it was
coming from the infinity. In this case all virtual photons (see (17)-(21),
Appendix 2) will be reflected from the both sides of the mirrors (in the
region of frequencies where the mirrors reflection coefficient is high).
The mirrors in this case can be done from thin ($\sim 10 \mu m$) Aluminum
foil transparent for the electron beam.

The broadband radiation is possible in a similar system if we will use
the system of mirrors arranged at different (irregular) distances.

The transition radiation sources considered above can be based on the linear
accelerators.

In the case of Smith-Purcell radiation the relativistic particle is moving
along a row of mirrors at some distance $a$ from them. In this case only
small part of Coulomb fields of the particle (virtual photons) will enter
the gap between mirrors (to the length $\leq \lambda _u$) and will be
converted in the real photons. The energy of the Smith-Purcell radiation of
one particle will be much less then the energy of the transition radiation
shown in the Fig.7. The electron beam in this case will not be disturbed by
the mirrors and hence such system in principle can be installed in the straight
section of the storage ring. A system of mirrors with hole of radius $a$
arranged in the form of Fig.7 can be used as well.
     \vskip -1mm
\empty \hspace*{-6cm}
\begin{center}
\unitlength=1.00mm
\linethickness{0.4pt}
\begin{picture}(140.00,147.33)
\put(30.00,75.00){\vector(1,0){110.00}}
\put(30.00,60.00){\line(0,1){30.00}}
\put(40.00,60.00){\line(0,1){30.00}}
\put(36.00,70.00){\makebox(0,0)[cc]{\small $O$}}
\put(42.00,55.67){\makebox(0,0)[cc]{\small $M1$}}
\put(50.00,60.00){\line(0,1){30.00}}
\put(52.00,55.67){\makebox(0,0)[cc]{\small $M2$}}
\put(60.00,60.00){\line(0,1){30.00}}
\put(62.00,55.67){\makebox(0,0)[cc]{\small $M3$}}
\put(70.00,60.00){\line(0,1){30.00}}
\put(72.00,55.67){\makebox(0,0)[cc]{\small $M4$}}
\put(80.00,60.00){\line(0,1){30.00}}
\put(82.00,55.67){\makebox(0,0)[cc]{\small $M5$}}
\put(90.00,60.00){\line(0,1){30.00}}
\put(92.00,55.67){\makebox(0,0)[cc]{\small $M6$}}
\put(100.00,60.00){\line(0,1){30.00}}
\put(102.00,55.67){\makebox(0,0)[cc]{\small $M7$}}
\put(110.00,60.00){\line(0,1){30.00}}
\put(112.00,55.67){\makebox(0,0)[cc]{\small $M8$}}
\put(30.00,90.00){\line(3,2){86.00}}
\put(116.00,147.33){\line(0,-1){144.67}}
\put(116.00,2.67){\line(-3,2){85.67}}
\put(64.33,118.00){\makebox(0,0)[cc]{\small $M^{conic}$}}
\put(140.00,70.00){\makebox(0,0)[cc]{\small $z$}}
\put(75.00,-10){\makebox(0,0)[cc]{\small Fig.7: The scheme of the transition
radiation of the electrons passing through the system }}
\put(80.00,-15.67){\makebox(0,0)[cc]{\small of 8 plane mirrors $M1 - M8$. The
emitted radiation is gathered by a conic mirror $M^{conic}$}}
\put(55.00,0.67){\makebox(0,0)[cc]{\small }}
\end{picture}
\end{center} \vskip 20mm

We have investigated the case of the particle radiation in a system of plane
parallel circular mirrors. Another systems can be used. For example a system
of the rectangular waveguides closed from one side and opened from another
side will permit to make more compact source of radiation. The electron beam
will penetrate these waveguides in turn in the direction perpendicular to
theirs walls. The radiation from the opened sides of the waveguides can be
gathered by a long plane mirror. Another example can be a system of pairs of
small plane mirrors turned at the angle 45$^o$ like in Fig.5 with separation
lenses and a long plane mirror to gather the emitted radiation.

\subsection{Stimulated transition radiation.}

If the system of two parallel circular mirrors shown on the Fig.6 will be
closed by a cylinder mirror then the transition radiation emitted on the
mirrors $M1$ and $M2$ will be reflected by the cylinder mirror and will be
returned to the central region of the system with the longitudinal component
of the electric field strength. Closed cylindrical cavity (resonator) will
be formed. Such cavity can be exited by a train of bunches to high value
electric field strength and high power can be extracted from this cavity
through a special hole or semitransparent window.

Usually at the center of such cavity a hole is produced for the electron
beam. The quality of the cavity with a hole is high for the wavelengths much
higher then the diameter of the hole. A system of such cavities in the form
of the iris-loaded waveguide is used as well.

In closed resonators the infinitely high amount of the longitudinal modes
exist. They can be exited by the electron beam with high efficiency if the
electron bunches of the beam will enter the resonator in the decelerating
phases and if the wavelengths of these modes of the order of the electron
bunch length and higher. Such resonator systems will emit electromagnetic
radiation with line spectrum.

\section{The long wavelength sources based on the storage rings.}

In the Fig.8 a principle scheme of an electromagnetic source based on a
storage ring is presented. The storage ring includes 4 banding magnets and
straight sections. Modern storage rings have more complicated magnetic
structure and can include many banding magnets quadrupole lenses and another
control systems.
              \vskip 18mm
\begin{center}
\unitlength=1.00mm
\linethickness{0.4pt}
\begin{picture}(84.34,55.00)
\put(35.00,42.50){\oval(30.00,25.00)[]}
\put(35.00,42.50){\oval(26.00,21.00)[]}
\put(52.33,42.00){\makebox(0,0)[cc]{{\tiny SS}}}
\put(50.67,31.33){\makebox(0,0)[cc]{{\tiny BM}}}
\put(15.67,54.00){\vector(1,0){68.67}}
\put(35.00,54.00){\circle*{0.80}}
\put(70.00,54.00){\circle*{0.80}}
\put(70.62,51.33){\makebox(0,0)[cc]{\tiny A}}
\put(83.67,51.00){\makebox(0,0)[cc]{\tiny z}}
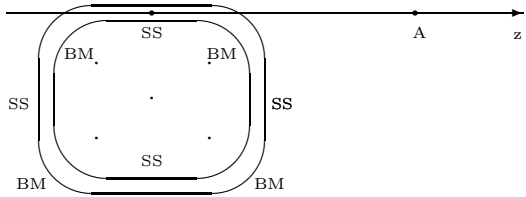
\put(35.00,20.00){\makebox(0,0)[cc]{{\small Fig.8: The scheme of the
long wavelength source based on a storage ring.}}}
\put(46.00,15.00){\makebox(0,0)[cc]{{\small BM are the banding magnets,
SS are the straight sections of the }}}
\put(45.00,10.00){\makebox(0,0)[cc]{{\small storage ring. A is the
observation point.}}}
\put(52.33,42.00){\makebox(0,0)[cc]{{\tiny SS}}}
\put(17.33,42.00){\makebox(0,0)[cc]{{\tiny SS}}}
\put(35.00,51.33){\makebox(0,0)[cc]{{\tiny SS}}}
\put(19.00,31.33){\makebox(0,0)[cc]{{\tiny BM}}}
\put(25.33,48.67){\makebox(0,0)[cc]{{\tiny BM}}}
\put(44.33,48.67){\makebox(0,0)[cc]{{\tiny BM}}}
\put(42.67,47.33){\circle*{0.50}}
\put(27.67,47.33){\circle*{0.50}}
\put(27.67,37.33){\circle*{0.50}}
\put(42.67,37.33){\circle*{0.50}}
\put(35.00,42.67){\circle*{0.50}}
\put(35.00,34.33){\makebox(0,0)[cc]{{\tiny SS}}}
\end{picture}
\end{center}
\vskip 8mm

In all storage rings the main part of the energy is emitted from the banding
magnets in the form of synchrotron radiation propagates in a small
vertical range of the angles and in near 2$\pi $ horizontal range of angles.
Only in a small ($\sim 10/\gamma $) horizontal range of angles near the
direction of the axis of the straight section the emitted radiation will have
the properties which are differ from the properties of the synchrotron
radiation \cite{bes1b}. Insertion devices can be installed in the straight
sections of the storage rings to increase the intensity to shift the hardness
to change the shape of the spectrum and to tune the polarization of the
emitted radiation. Usually such devices are the undulators of different types
\cite{bes4}, \cite{bes5} or different banding magnets creating some desirable
shapes of the magnetic field along the axes of the straight sections of the
storage rings \cite{bes1b}.

The spectral-angular density of the energy of the long wavelength synchrotron
radiation $\lambda > \lambda _c$ emitted from the bending magnet of the
radius $\rho $ in the range of angles $\vartheta \ll \vartheta _c$ can be
presented in the form
\cite{jackson}
             \begin{equation}  
             {\partial ^2\varepsilon \over \partial {\omega }\partial o} =
           {3e^2\gamma ^2\Gamma ^2({2\over 3})\over 2^{2/3}\pi ^2c}({\omega
           \over \omega _c})^{2/3}[1 + {\Gamma ^2({1\over 3})\over 2^{2/3}
           \Gamma ^2({2\over 3})}\vartheta ^2({\omega \over \omega _c
           })^{2/3}],    \end{equation}
where $\omega _c = 3\omega _0\gamma ^3$, $\omega _0 =c/\rho = eB_m/mc\gamma $,
$\vartheta _c = (\omega _c/\omega)^{1/3}$, $\Gamma ({1\over 3}) \simeq 2.68$,
$\Gamma ({2\over 3}) \simeq 1.35$. Inside the range of angles $\theta \leq
\vartheta _c/\gamma $ the main part of the energy of the synchrotron
radiation will be emitted.

It follows from the equation (28) that the spectral-angular density of the
synchrotron radiation increases with the energy of particles ($\sim \gamma
^2$) and with the increasing of the banding radius $\rho $ or the same with
the decreasing of the value of the magnetic field.

In the case of limited particle trajectories with straight sections which
takes place in the storage rings the spectrum of radiation emitted in the
directions near to the direction of the axis of a straight section (under the
angles $\sim 1/\gamma$ to this direction) is shifted in the long wavelength
region. The value of this shift can be increased at low frequencies by using
optical lenses, concave mirrors and other elements to select light emitted
from the end part of the electron trajectory in the banding magnet
\cite{bes1a}, \cite{bes1b}. Preliminary experiments confirmed the possibility
of such selection of the radiation emitted by electrons in the fringed fields
of the synchrotrons and storage rings \cite{al3}

The spectral-angular density of the energy of the long wavelength ($\lambda >
d/2\gamma ^2$ electromagnetic radiation emitted in the direction near to the
direction of the straight section of the storage ring without using of the
selective elements can be presented in the form
             \begin{equation}  
             {\partial ^2\varepsilon \over \partial {\omega }\partial o} =
             {4e^2\gamma ^2\over \pi ^2c}{\vartheta ^2\over (1 + \vartheta
             ^2)^2}\sin ^2{\omega \over \omega _l},    \end{equation}
where $\omega _l = 4\gamma ^2c/l(1 + \vartheta ^2)$, $l$ is the length of
the straight section of the storage ring, $d$ is the gap of the banding magnet
of the storage ring \cite{bes1b}. This value does not depend on the magnetic
field strength. It has a maximum at the wavelengths $\lambda = 2l/n\gamma ^2$,
and at the angle $\vartheta = 1$, where $n$ = 1, 2, 3... is the integer.

The spectral-angular density of the energy of the electromagnetic radiation
emitted  in these directions at the wavelengths $\lambda \sim \lambda _l$
can be increased if three banding magnets will be installed in the straight
section of the storage ring. First and third banding magnets must have half
of the optimal magnetic fields (11) and second banding magnet must have the
optimal magnetic field of the opposite polarity. Linear polarized radiation
will be emitted from such system. Three bending magnets of such configuration
does not disturb the electron orbits in the storage ring at arbitrary
magnetic fields when the ratio $1/2$ is fulfilled. The increasing of the value
of the magnetic field strengths in such system to the values much higher then
optimal will lead to the case of enhanced synchrotron radiation if the value
of the magnetic field strengths in the magnets will be less then the value
of the magnetic field strength in the bending magnets of the storage ring
\cite{art1}, \cite{art2}, \cite{art3}.

The circular polarized long wavelength radiation in such geometry can be
realized when two triplet systems of such kind will be used. The direction
of the magnetic field of the second triplet must be perpendicular to the
direction of the first one and the second triplet must be shifted on the
distance of a half distance between the first and second magnets of the
first triplet \cite{bes1b}.

{\it Example.} The storage ring of the Tohoku University will have the energy
of 1.5 GeV, the magnetic field strength in the banding magnets $1.25 \cdot
10^4$ Gs, the natural bunch length $4.3$mm, emittance $\epsilon \sim 7.3$
nm-rad, momentum spread $6.6 \cdot 10^{-4}$, circumference $C = 187$ m, 12
dispersion-free straight sections of 5 m long and 2 straight sections of 12 m
long, betatron tunes $\nu _{x}$ = 12.2 and $\nu _z$ = 3.15. The transverse
beam dimensions evaluated by the expression $\sigma _{x,z}= \sqrt
{C\epsilon/2\pi \nu _{x,z}}$ are equal: $\sigma _x = 2.6 \cdot 10^{-2}$ mm,
$\sigma _z = 1.3 \cdot 10^{-2}$ mm.

In this case the maximum of the long wavelength radiation will be emitted at
the angle $\vartheta = 1$ to the axis of the long straight section: $\lambda
_l \sim 1.3 \cdot 10^{-3}$ mm.

\section{On the choice of the scheme of the long wavelength broadband source
of coherent radiation}

The problem of the choice of the optimal scheme of the electromagnetic
radiation
source based on the electron beams is determined by the real conditions.
If you need some source and have no accelerator then you will have one
solution. If you need the same source and have one ore more accelerators then
you can choose another decision of the problem.

The accelerator is the most expensive and large part of the sources. If you
have accelerator then you will try to adapt it to the solution of the source
problem. Your existing accelerators probably were used and are used now
in other fields of science or technology. Their parameters like electron
beam energy, current, transverse and longitudinal dimensions, angular and
energy spreads (emittances) can be far from optimal ones for the solution of
your problem. Nevertheless the non-optimal solution with non-brelliant
parameters can suit you.

At present the optimal solution of the problem of the electromagnetic
radiation sources based on electron (or probably ion) beams is well known.
It depends on the wavelength region, power, monochromaticity, directivity
and so on.

If you need high power, high brightness source of the quasi-monochromatic
optical to
hard X-Ray (soft $\gamma -$Ray) regions then you ought to use the
undulator radiation sources based on the storage rings \cite{alf1},
\cite{bes6}. If you need high average intensity source of continuous hard
quasi-monochromatic $\gamma -$Rays with tuned polarization then you ought to
use the backward Compton scattering source based on a laser and on a storage
ring. Backward Rayleigh scattering source can be used in future when the
storage rings with relativistic ions (like LHC, RHIC) will appear \cite{kim}.

If you need high power monochromatic source of coherent radiation in mm-to
vacuum ultraviolet regions you ought to use prebunched or ordinary (with
self-bunching) free-electron lasers based on storage rings and linear
accelerators.

The spontaneous incoherent undulator radiation sources and free-electron
lasers use undulators. Undulators are the devices which force the
particles to move along the periodical trajectory (sine, helix and so on).
The properties of the radiation are determined by the trajectory. The same
trajectories can be produced by different undulators. For example electrons
are moving along the helical trajectories in homogeneous magnetic field, in
helical undulator, in the field of the electromagnetic wave and so on.
Crystals can be considered as a natural undulators. Undulators based on the
magnets and electromagnetic waves of lasers does not disturb the electron
beam and are used in the sources in mm-to $\gamma -$ray regions.

If you need the broadband source of the optical to X-Ray regions then you
can use the synchrotron radiation \cite{winick} source or the undulator
radiation source based on the undulator forming the magnetic field by a
definite low \cite{bes4} or the system of magnets with alternating polarity
irregular arranged along some axis. Usually the intensity of the spontaneous
incoherent radiation of such sources is enough for users.

If you need the broadband or the quasi-monochromatic source of the $\gamma
-$ray regions then you can use the breamsstrahlung radiation produced by the
electron beams in the amorphous media and coherent breamsstrahlung or
channeling radiation in crystals.

The intensity of the broadband sources is decreased with the increasing
of the emitted wavelength. The problem of the long wavelength high intensity
broadband sources appear. This problem is aggravated by the condition of the
broadband sources which means that you can not use the coherent radiation of
the microbunches arranged on short distances. The distance between
microbunches must be much larger then the length of the wavepackets emitted by
one microbunch in the magnet system because the overlapping of the broadband
wavepackets can lead to the decreasing of the emitted energy (in the case of
the undulator radiation sources the distance between microbunches can be
$\lambda $ that is $K$ times shorter, then the length of the wavepackets,
electric field strength of the wavepackets overlapped in phase can be $K$
times higher and the intensity $K^2$ times higher). The using of the
periodical train of the electron bunches will lead to the line structure of
the emitted spectra (see (7)). Only the coherent radiation of the individual
arranged at large distances bunches is possible in this case. Some schemes of
such kind were considered in this paper. Non periodical train of the electron
bunches can permit to decrease the average distance between electron bunches
and this way to help to push forward the problem under the consideration.

The ability of some schemes of the particle radiation in external fields and
in media are presented in the Appendix 3. We can see that the most efficient
schemes of the broadband particle radiation sources in mm-to soft IR region
could be sources based on Cherenkov radiation in the transparent media and on
the transition radiation system based on a system of parallel mirrors
(Fig.7). Unfortunately the transparent media will be ionized and destroyed by
the electron beam and the electron beam in turn will be disturbed by
scattering process in media. That is why the Cherenkov radiation schemes are
not used in the sources under consideration. The transition radiation
sources using thin metal foils can be based on linear accelerators. Using the
transparent media and mirrors with holes does not solve this problem as
sources based on such systems permit generate effectively the electromagnetic
radiation on the wavelengths of the order of the aperture of such holes and
longer. The spectral intensity of the  radiation emitted at a given frequency
in this case does not depend on the electron energy and only lead to the
extension of the bandwidth of the emitted radiation. The most efficient
scheme of the long wavelength and broadband source of coherent radiation
apparently is the scheme of the prebunched free-electron laser based on an
undulator and on an opened resonator.

                  \section{Conclusion.}

Different  schemes of the broadband sources of coherent radiation in the
millimeter and shorter wavelength regions based on bunched particle beams
were considered in this paper. The most efficient scheme of such source is
apparently the prebunched free-electron laser based on an undulator and on an
opened resonator.

In order to produce the long wavelength radiation in these regions on
high energy accelerators ($\sim$ 100 MeV) the long period undulators with
high deflecting parameters can be used. In order to generate the broadband
radiation it is necessary  to use two or several banding magnets of the
alternating polarity with high deflecting parameters and placed on different
distances one after another (to break the periodicity).

Notice that the prebunched free-electron lasers based on undulators or
similar systems and opened resonators have an important feature. The emitted
energy in such lasers is extracted from a hole of small diameter (much
smaller then the mirror diameter). It will lead to the increased brightness
of such sources.

The authors thanks Professor M.Ikezawa for the formulation of the problem
of the powerful long wavelength sources of the continuous broadband radiation
and for helpful discussions.

\vfill\eject

           \section{Appendix 1.}

When the angular and energy spreads of the electron beam are small
          \begin{equation} 
        {\Delta \varepsilon \over \varepsilon} \ll {1 \over K}, \hskip 15mm
        \Delta \theta \ll \frac {1}{\gamma K}
          \end{equation}
and when the influence of the electromagnetic field of the wavepackets on the
electron trajectories is small then all electron trajectories of the beam
will be similar. In this case the wavepackets emitted by each electron will be
similar as well and will be scattered in time. They will be described by the
expression $\vec E_i(t - t_i)$ of the form (13) if we accept $N_b = 1$ and
remark $t_w$ on $t_i$, where $i$ is the electron number. Then the total field
strength of the stored wavepacket will be equal
          \begin {equation} 
         \vec E(t) = \int _{t_1}^t\dot N_{\gamma }(t_i)\vec E(t - t_i)dt_i,
                   \end{equation}
where $\dot N_{\gamma } = dN_{\gamma }/dt$ is the flux of the electron
wavepackets trough the point of observation at the moment $t_i$, and $t_1$ is
the moment of arrival of the first wavepacket at the point of observation.
The value $\dot N_{\gamma}(t_i) = \dot N_e(t_i) = i(t_i)^{`}/e$, $i(t)$ is the
beam current, $t^{`}_i = t_i - (y - y_o)/c$ is the moment of the electron
entering the undulator in position $y_o$.

The Fourrier's transform of this equation is
          \begin {equation} 
         \vec E_{\omega} = 2\pi \vec E_{\omega 1}\dot N_{e\omega } =
         2\pi \vec E_{\omega 1}i_{\omega}/e,
                   \end{equation}
where $\vec E_{\omega 1}$ relates to the first particle of the bunch and the
moment $t_i|_{i = 1} = 0$.

To derive the equation (32) we used the expression
       $${1\over 2\pi}\int _{-\infty}^{+\infty}\vec E(t - t_i)exp (i\omega t)
         dt = \vec E_{\omega 1}\exp ^{i\omega t_i}, $$

In the case of the extended electron bunches the spectrum will be determined
by the product of the spectral function of the radiation emitted by the
"pointlike bunch" consisting on one electron and the spectral function 2$\pi
i_{\omega }/e \leq N_b$ of the bunch current $i(t)$

The Fourrier's component of the electric field strength of the wave emitted by
point-like charge (13) is
         \begin{equation} 
       \vec E_{\omega } = \vec E_{\perp }e^{-r^2/\sigma
_{\gamma}^2}\vec \psi _{\omega } e^{i\omega t_w},
\end{equation}
where $\vec \psi _{\omega } = \vec e\psi _{\omega -
\omega _1} + \vec e^*\psi _{\omega + \omega _1}$, $\vec e^* = \vec e_x
+ i\vec e_z$, $\tau = t - t_w$,
         $$\psi _{\omega \pm \omega _1} =
         {1\over 2\pi}\int _0^{T}f(\tau) e^{i\omega \tau }d\tau =
         {T\over 2\pi}{\sin {(\omega \pm \omega _1)T \over 2} \over
         {(\omega \pm \omega _1)T\over 2}}e^{i(\omega \pm \omega
         _1)T/2}.$$

When the undulator have the whole number of the
periods $K$ then $T = KT_1$ and $f_{\omega \pm \omega _1} = \omega _{1}
^{-1}(\sin K\sigma _{1 \pm}/ \sigma _{1 \pm})\exp (iK\sigma _{1 \pm})$,
$\sigma _{1\pm} = \pi (\omega \pm \omega _1)/\omega _1$.  \vfill\eject

           \section{Appendix 2.}

The spectrum of virtual quanta (energy per unit area $\partial S$ per unit
frequency interval $\partial \omega$) for the ultrarelativistic electrons
given by \cite{jackson}
                      \begin{equation}   
           {dI\over d\omega } = {\partial ^2\varepsilon ^{eq}\over \partial
           \omega \partial S} = {e^2\over \pi ^2c\beta ^2b^2}({\omega b\over
           v\gamma })^2[K_1^2({\omega b\over v\gamma}) + {1\over \gamma}
           K_0^2({\omega b\over          v\gamma})],           \end{equation}
where $b$ is the impact parameter (in our case the distance from an electron
trajectory to the given point on the surface of the mirror), $\beta =
v/c$, $v$ is the particle velocity, $\gamma = \varepsilon /mc^2$
is the relativistic factor, $\omega = 2\pi c/\lambda $, $\lambda $ is the
wavelength of the emitted radiation, $K_i$ is a modified Bessel function
of order $i$. It corresponds to the case of wavepacket of real electromagnetic
waves with the time dependence of the transverse electric and magnetic field
strengths components and the longitudinal electrical field strength along the
axis $x_2$, $x_3$ and $x_1$ accordingly of the form \cite{jackson}
                      \begin{equation}   
           E_{x_2} = B_{x_3} = {eb\gamma \over (b^2 + \gamma ^2 v^2t^2)^{3/2}},
    \hskip5mm   E_{x_1} = {e\gamma ct\over (b^2 + \gamma ^2 v^2t^2)^{3/2}}.
           \end{equation}

In the limit $\gamma \to \infty $ the values $x = \omega b/v\gamma \ll 1$,
$K_0(x) \simeq 0.5772 - ln(x/2)$, $K_1(x) \simeq 1/x$ and hence the spectral
density of the virtual quanta at a given frequency and impact parameter does
not depend on $\gamma $ and tends to it's maximum value
                      \begin{equation}   
           {\partial ^2\varepsilon ^{eq}\over \partial \omega \partial S} =
        {e^2\over \pi ^2c\beta ^2b^2},           \end{equation}
The increasing of the electron energy in this case lead to the expansion of
the spectral bandwidth of the virtual quanta in the high frequency region
and total density of the energy of these quanta.

In collision problems the frequency spectra (1) must be summed over various
possible impact parameters to get the energy per unit frequency interval
present in the equivalent radiation field. It has the form
                      \begin{equation}   
           {\partial \varepsilon ^{eq}\over \partial \omega } =
        {2e^2\over \pi c}\{xK_0(x)K_1(x) - {x^2\over 2}[K_1^2(x) - K_0^2(x)]\},
           \end{equation}
where $x = \omega b_{min}/v\gamma $, $b_{min}$ is a minimum impact parameter
for a given problem (in our case $b_{min}$ is the radius $a$ of the hole in
the mirror with circular aperture).

In the ultrarelativistic case $\beta \sim 1$, $\gamma \gg 1$ for low
frequencies $x\ll 1$ or $\omega \ll \omega _c = \gamma c/b_{min}$
the energy per unit frequency interval reduces to
                      \begin{equation}   
           {\partial \varepsilon ^{eq}\over \partial \omega } =
    {2e^2\over \pi c}[\ln ({1.123\omega _c\over \omega }) - {1\over 2}],
           \end{equation}
whereas for high frequencies $\omega \gg \omega _c$
                      \begin{equation}   
           {\partial \varepsilon ^{eq}\over \partial \omega } =
        {e^2\over 2c} \exp ^{-2\omega /\omega _c}.
           \end{equation}

The number of the virtual quanta per one electron per frequency interval
$\Delta \omega $ converted to the real photons by the mirror
                      \begin{equation}   
        N_{\gamma } = {1\over \hbar \omega }{\partial \varepsilon
^{eq}\over \partial \omega }{\Delta \omega } \simeq {2\alpha
\over \pi}\ln{1.123\omega _c\over \omega }{\Delta \omega \over \omega}.
           \end{equation}

\vfill\eject

                     \section{Appendix 3.}

To compare the ability of different schemes of the particle radiation in
external fields and in media below we will present spectral-angular density
of photons $\partial ^2N_{\gamma }/\partial \omega \partial o = (1/\hbar
\omega) (\partial ^2\varepsilon /\partial \omega \partial o)$ in an
electromagnetic wave emitted by one particle in the corresponding fields or
media. The coherent radiation of the particle bunches can be calculated by
introducing the coherence factor which is near the same for all of these
schemes and hence the comparison will be valid for the coherent radiation
sources as well.

1. An electron comes from the infinity with the initial velocity $\vec v_1$
directed at the angle $\alpha < 0$ to the axis $z$. Then the electron
is deflected by a magnet at the angle $2\alpha $ in the $xz$ plane and go
out with the final velocity $\vec v_2$.

In this case the spectral-angular density of photons is determined by
expressions (4), (6). It takes on maximum value
               \begin{equation} 
               {\partial ^2N_{\gamma }\over \partial \omega
               \partial o}|_{\theta = 0} = {4\alpha \gamma ^2\over \pi ^2
               \omega }. \end{equation}
at the banding angle $2\alpha = 2/\gamma $ and in the direction of the axis
$z$ where the strange parameter (6) is maxima \cite{bes1a}.

The total number of photons emitted in the relative range of frequencies
$\Delta \omega /\omega$ in this case
               \begin{equation} 
               \Delta N_{\gamma } = {4\alpha \ln \gamma \over \pi }{\Delta
               \omega \over \omega}. \end{equation}

This case can not be realized as suppose the infinite electron trajectories.
Radiation emitted in such magnet installed in the storage ring and separated
by focusing elements could lead to the values which are near to (41), (42).

2. An electron is moving in the homogeneous magnetic field along the circular
orbit of a radius $\rho $ and emit synchrotron radiation.

In this case according to (28)

           $$ {\partial ^2N_{\gamma } \over \partial {\omega }\partial o} =
           {3\alpha \gamma ^2\Gamma ^2({2\over 3})\over 2^{2/3}\pi ^2\omega}
           ({\omega \over \omega _c})^{2/3}[1 + {\Gamma ^2({1\over 3})\over
           2^{2/3} \Gamma ^2({2\over 3})}({\omega \over \omega_c})^{2/3}
           \vartheta ^2], $$
           \begin{equation}  
            \Delta N_{\gamma } = {3\alpha \gamma ^2\Gamma ^2({2\over 3})\over
            2^{2/3}\pi ^2}({\omega \over \omega _c})^{2/3}[1 + {\Gamma
            ^2({1\over 3})\over 2^{2/3} \Gamma ^2({2\over 3})}({\omega \over
            \omega_c})^{2/3}\vartheta ^2]{\Delta \omega \over \omega}\Delta o,
            \end{equation}
where $\Delta o$ is the solid angle ($\Delta \varphi \leq 2\pi, \Delta \theta
\leq \theta _c$) which can be accepted by the experimental installation.

The values (42) are proportional to $\omega _c^{-2/3}$ or $B_m^{-2/3}$. We
can increase these values if will use the long banding magnets with weak
magnetic fields (11) installed in the straight section of the storage ring as
was discussed in the section 6.

3. An electron emit radiation in a helical undulator. The period of the
undulator is $\lambda _u$, the number of periods is $K$ and the deflecting
parameter is $p_{\perp}$.

In this case the frequency of the undulator radiation emitted on the first
harmonic along the axis $z$ will be determined by the expression (12) and
the number of photons emitted in the frequency range $\Delta \omega /\omega$
             \begin{equation}  
            \Delta N_{\gamma 1} = (2\pi /3)\alpha Kp_{\perp}^2f(\omega/\omega
            _m)(\Delta \omega /\omega),                 \end{equation}
where $\omega _{1m} = 2\pi c/\lambda _1(\vartheta = 0), f(\xi) = 3\xi (1 -
2\xi + 2\xi ^2)$.

4. Prebunched free-electron laser based on an undulator and on an opened
resonator.

The number of photons emitted by one electron in the prebunched free-electron
laser based on a helical undulator and on an opened resonator when electron
bunch consist of one electron according to (14) is equal
     \begin{equation}   
       \Delta N_{\gamma}^{out} = 8\alpha Q{K\lambda _u\over l_R}
       {p_{\perp }^2\over 1 + p_{\perp}^2},              \end{equation}
where $K\lambda _u/l_R < 1$.

5. An electron emit transition radiation when crossing the plain mirror.

In this case the spectral-angular density of photons and the total number of
photons emitted in the relative range of frequencies $\Delta \omega /\omega$
is determined by expressions (17), (21)
               \begin{equation} 
               {\partial ^2N_{\gamma }\over \partial \omega \partial o} =
               {\alpha \gamma ^2\over \pi ^2 \omega }{\vartheta ^2\over 1 +
               \vartheta ^2}, \hskip 1cm
               \Delta N_{\gamma } = {2\alpha \over \pi }[\ln 2\gamma -1]
               {\Delta \omega \over \omega}. \end{equation}

The using a system of two or $M > 2$ plane mirrors will permit to reach 2M
times more high values of the spectral-angular density and total number of
the emitted photons (see section 5.3).

6. An electron emit Cherenkov radiation when it pass through the transparent
media.

The spectral distribution of the Cherenkov radiation emitted by an electron
in a media and the total number of photons emitted in the relative range of
frequencies $\Delta \omega /\omega$ are determined by the expressions
\cite{jackson}.
               \begin{equation} 
               {\partial N_{\gamma }\over \partial \omega} = {\alpha l\over c}
               (1 - {1\over \epsilon \beta ^2}), \hskip 1cm
               \Delta N_{\gamma } = {2\pi \alpha l\over \lambda }(1 -
               {1\over \epsilon
               \beta ^2}){\Delta \omega \over \omega}, \end{equation}
where $\epsilon $ is the dielectric constant, $l$ is the length of the
electron  trajectory in media.

It follows from the expression (47) that the number of the emitted photons
does not depend on $\gamma $ when $\gamma ^2 \gg \epsilon /(1 - \epsilon)$
and tends to maximum when $\gamma \to \infty $.

\vfill\eject

\newpage \empty \vspace * {15cm}
\footskip 10cm 
\begin{center} Signed to print October 15, 1996

Order No 186. 50 copies printed. P.l. 1.8

--------------------------------------

Printed in RIIS FIAN.

Moscow, V-333, Leninsky prospect, 53.
\end{center}

\end{document}